\documentclass[graybox]{svmult}
\usepackage{mathptmx}       % selects Times Roman as basic font
\usepackage{helvet}         % selects Helvetica as sans-serif font
\usepackage{courier}        % selects Courier as typewriter font
\usepackage{type1cm}        % activate if the above 3 fonts are
\usepackage{makeidx}         % allows index generation
\usepackage{graphicx}        % standard LaTeX graphics tool
\usepackage{multicol}        % used for the two-column index
\usepackage[bottom]{footmisc}% places footnotes at page bottom

\usepackage{amsmath, amsfonts, amssymb, amscd, mathrsfs}
\usepackage[initials,nobysame,short-months]{amsrefs}

\usepackage{manfnt}          % für die Todo boxen

\smartqed                    % setzt qed and das Zeilenende

\makeindex             % used for the subject index
\def\d{ \mathrm{d} }								% integral d
\def\I{ \mathrm{i} }								% imaginary unit
\def\E{ \mathrm{e} }								% e
\def\T{ \mathrm{T} }								% T - transpose

\def\NN{ \mathbb{N} }								% set of integers
\def\ZN{ \mathbb{Z} }								% set of whole numbers
\def\CN{ \mathbb{C} }								% complex numbers
\def\RN{ \mathbb{R} }								% real numbers

\def\B{ \mathcal{B} }								% Bernstein space
\def\PW{ \mathcal{PW} }								% Paley Wiener spaces
\newcommand{\Op}[1]{\mathrm{#1}}						% Operatoren - general
\DeclareMathOperator*{\esssup}{ess\,sup}				% ess sup
\DeclareMathOperator*{\cls}{\overline{span}}			     % closed linear span

\begin{document}

\title*{System Approximations and Generalized Measurements in Modern Sampling Theory}
\author{Holger Boche and Volker Pohl}
\institute{
Holger Boche, Volker Pohl
\at Technische Universit{\"a}t M{\"u}nchen, Lehrstuhl f{\"u}r Theoretische Informationstechnik, Arcisstra{\ss}e 21, 80333 M{\"u}nchen, Germany,
\email{ {boche, volker.pohl}@tum.de}
\and
This is a preprint of an invited chapter which appears in \emph{Sampling Theory - a Renaissance}, Editor: \emph{G. Pfander}, Springer-Birkh{\"a}user, 2015.}
\maketitle

% ==============================
% ========== ABSTRACT ==========
% ==============================
% --- Online Version ---
\abstract*{Each chapter should be preceded by an abstract (10--15 lines long) that summarizes the content. The abstract will appear \textit{online} at \url{www.SpringerLink.com} and be available with unrestricted access. This allows unregistered users to read the abstract as a teaser for the complete chapter. As a general rule the abstracts will not appear in the printed version of your book unless it is the style of your particular book or that of the series to which your book belongs.
Please use the 'starred' version of the new Springer \texttt{abstract} command for typesetting the text of the online abstracts (cf. source file of this chapter template \texttt{abstract}) and include them with the source files of your manuscript. Use the plain \texttt{abstract} command if the abstract is also to appear in the printed version of the book.}

% --- Printed Version ---
\abstract{
This paper studies several aspects of signal reconstruction of sampled data in spaces of bandlimited functions.
In the first part, signal spaces are characterized in which the classical sampling series uniformly converge,
and we investigate whether adaptive recovery algorithms can yield uniform convergence in spaces where non-adaptive sampling series does not.
In particular, it is shown that the investigation of adaptive signal recovery algorithms needs completely new analytic tools since the methods used for non-adaptive reconstruction procedures, which are based on the celebrated Banach-Steinhaus theorem, are not applicable in the adaptive case.\\
The second part analyzes the approximation of the output of stable linear time-invariant (LTI) systems based on samples of the input signal,
and where the input is assumed to belong to the Paley-Wiener space of bandlimited functions with absolute integrable Fourier transform.
If the samples are acquired by point evaluations of the input signal $f$, then there exist stable LTI systems $\Op{H}$ such that the approximation process does not converge to the desired output $\Op{H} f$ even if the oversampling factor is arbitrarily large.
If one allows generalized measurements of the input signal, then the output of every stable LTI system can be uniformly approximated in terms of generalized measurements of the input signal.\\
The last section studies the situation where only the amplitudes of the signal samples are known.
It is shown that one can find specific measurement functionals such that signal recovery of bandlimited signals from amplitude measurement is possible, with an overall sampling rate of four times the Nyquist rate.
}

% ==================================
% ========== INDRODUCTION ==========
% ==================================
\section{Introduction}
\label{sec:Indro}

The great success of digital signal proceeding lies in the fact that analog signals observed in the physical world can equivalently be represented by a sequence of complex numbers.
These digital signals can then be processed and filtered very quickly and efficiently on digital computers.
Sampling theory is the theoretical foundation of the conversion from analog to digital signals and vice versa.
Because of its fundamental importance for modern information theory, signal processing and communications, there exists a long and extensive list of impressive research results in this area starting with the seminal work of Shannon \cite{Shannon_IRE1949}. We refer to excellent survey articles and textbooks such as \cites{Higgins_Sampling1, Jerri_ShannonSampThm_77,Seip,Unser_PocIEEE00,Zayed_ShannonSampTheorie} and to \cites{Butzer_RaabesWork_AA11,Butzer_JFAA11} for historical comments on the topic.

Sampling theory was originally formulated for bandlimited signals with finite energy.
Later these results were extended to non-bandlimited signals \cites{ButzerStens_SIAM92,Eldar_beyond_2009,EP_IEEE_SP10,Ferreira_SPL95} and to broader classes of bandlimited functions, in particular to functions which do not necessarily have finite energy \cites{Campbell_SIAM68,Pfaffelhuber_IEEEIT71,Zakai_65}.
But these results often took into consideration only the pointwise convergence of the reconstruction series.
From a practical point of view, however, it is often necessary to control the peak value of the reconstructed signal because electronic circuits and devices (like amplifiers, antennas, etc.) have only limited dynamic ranges. Moreover, the energy efficiency of these devices usually and largely depends on this dynamic range \cite{Wunder13}.
To control the peak value of the signals, one has to investigate the uniform convergence of the sampling series.
This will be done in some detail in the first part of this article (Sec.~\ref{sec:GUConv}).
The starting point will be a classical result \cite{Brown_67} which shows that the uniform Shannon sampling series is
locally uniformly convergent for all bandlimited signals with an absolute integrable Fourier transform.
Then we present several extensions of this result to larger signal spaces and we investigate whether it is possible to have global uniform convergence on the entire real axis.
In particular, we discuss the influence of the sampling points and we investigate whether its is possible to apply adaptive reconstruction algorithms to obtain signal recovery methods which are uniformly convergent.
Classical sampling series, like the one of Shannon, are fixed for the whole signal space under consideration. These series may or may not converge for all signals in this signal space.
But even if the sampling series does not converge for all functions in the space, it might be possible to adapt the reconstruction series to the actual signal to obtain a signal approximation which converges uniformly to the desired signal. However, it will be shown that for the common signal spaces of bandlimited signals, such an adaption of the recovery series essentially gives no improvement of the global uniform convergence behavior.
These investigations in Sec.~\ref{sec:GUConv} are strongly related to one of the cornerstones of functional analysis, namely to the theorem of Banach-Steinhaus\index{Theorem!of Banach-Steinhaus}.  
This important theorem is a very powerful and elegant tool to investigate non-adaptive algorithms, and in particular to prove the divergence of sampling series considered in this paper.
For adaptive algorithms, however, the Banach-Steinhaus theorem cannot be applied. Therefore completely new tools are necessary for the investigation of such recovery algorithms, which are related to some early works of Paul~Erd{\H o}s.

The second part of this article (Sec.~\ref{sec:SampBasedSP}) investigates another aspect of sampling based signal processing.
Whereas the sampling theorem deals with the reconstruction of a signal $f$ from its samples $\{f(\lambda_{n})\}$,
the main task of digital signal processing is often not to reconstruct $f$, but to process the sampled data $\{f(\lambda_{n})\}$ such that an approximation of the processed signal $g = \Op{H} f$ is obtained, where $\Op{H}$ is a certain linear transformation.
In other words, one wants a digital implementation of the analog system $\Op{H}$.
It seems to be widely accepted that such a digital implementation is always possible, at least for stable linear systems.
This is certainly true for bandlimited signals with finite energy.
However, for more general signal spaces, a digital implementation of stable LTI systems may fail, even on such spaces where the sampling series uniformly converges.
This remarkable observation is presented and discussed in Sec.~\ref{sec:SampBasedSP}.
Moreover, we also discuss the influence of data acquisition for possible signal based signal processing.
Usually, signals are sampled by point evaluations $f \mapsto \{f(\lambda_{n})\}_{n\in\ZN}$. However, more general measurement functionals are possible to digitize an analog signal.
It is shown that generalized measurement functionals will enable the digital implementation of analog signal processing schemes on spaces where a digital implementation based on point evaluations fail.

The last part (Sec.~\ref{sec:Phase Retrieval}) considers another application where generalized measurement functionals are necessary to guarantee signal recovery from signal samples.
Here we consider the situation where only the amplitude of the signal samples is available, but not the phase. This problem, known as \emph{phase retrieval}, plays an important role in many different applications.
Even though there is a long history of research \cite{Fienup1982_PhaseRetrieval} on phase retrieval, it is still not clear whether signal reconstruction from the amplitudes of point evaluations is in general, possible. Here we will show, however, that specifically designed measurement functionals will allow signal recovery of bandlimited functions from the knowledge of its amplitudes only. The recovery algorithm will be partially based on the sampling series investigated in Sec.~\ref{sec:GUConv}.

% =====================================================================
% ========== Notations ================================================
% =====================================================================
\section{Preliminaries and Signal Models}
\label{sec:Notations}
% Always give a unique label
% and use \ref{<label>} for cross-references
% and \cite{<label>} for bibliographic references
% use \sectionmark{}
% to alter or adjust the section heading in the running head

% ==========================================
\subsection{General Notations}

We use standard notations.
In particular, the set of all continuous functions on the real axis $\RN$ is denoted by $\mathcal{C}(\RN)$, and $\mathcal{C}_{0}(\RN)$ stands for all $f\in \mathcal{C}(\RN)$ which vanish at infinity.
Both spaces are equipped with the supremum norm $\|f\|_{\infty} = \sup_{t\in\RN} |f(t)|$.
If $1 \leq p < \infty$ or $p = \infty$, then
\begin{equation*}
	\|f\|_{p} = \left(\int^{\infty}_{-\infty} |f(t)|^{p}\, \d t \right)^{1/p}
	\qquad\text{and}\qquad
	\|f\|_{\infty} = \esssup_{t\in\RN} |f(t)|
\end{equation*}
is the $L^{p}$-norm of $f$, and $L^{p}(\RN)$ stands for the Banach space of all measurable functions on $\RN$ with finite $L^{p}$-norm.
Similarly, if $\mathbb{S} \subset \RN$ is a finite interval of length $|\mathbb{S}|$ on $\RN$, then $L^{p}(\mathbb{S})$ with $1 \leq p \leq \infty$ stands for the set of measureable functions on $\mathbb{S}$ with finite norm, defined by
\begin{equation*}
	\|f\|_{p} = \left(\frac{1}{|\mathbb{S}|}\int_{\mathbb{S}} |f(t)|^{p}\, \d t \right)^{1/p}
	\qquad\text{and}\qquad
	\|f\|_{\infty} = \frac{1}{|\mathbb{S}|}\esssup_{t\in\mathbb{S}} |f(t)|\;.
\end{equation*}
In particular, $L^{2}(\RN)$ and $L^{2}(\mathbb{S})$ are Hilbert spaces with the inner products
\begin{equation}
\label{equ:InnerProductL2}
	\left\langle f,g \right\rangle = \int_{\RN} f(t)\, \overline{g(t)}\, \d t
	\qquad\text{and}\qquad
	\left\langle f,g \right\rangle = \frac{1}{|\mathbb{S}|}\int_{\mathbb{S}} f(t)\, \overline{g(t)}\, \d t\;,
\end{equation}
respectively.
For any $f \in L^{1}(\RN)$, its \emph{Fourier transform}\index{Fourier transform} is defined by
\begin{equation*}
	\widehat{f}(\omega)
	= (\mathcal{F} f)(\omega) = \int^{\infty}_{-\infty} f(t)\, \E^{-\I t\omega}\, \d t\;,
	\qquad\omega\in\RN\;.
\end{equation*}
Because $L^{1}(\RN) \cap L^{2}(\RN)$ is a dense subset of $L^{2}(\RN)$, \emph{Plancherel's theorem} extends $\mathcal{F}$ to a unitary operator on $L^{2}(\RN)$.
There, $\mathcal{F}$ satisfies \emph{Parseval's formula} $\big\langle \widehat{f} , \widehat{g} \big\rangle = 2\pi \big\langle f, g \big\rangle$ for all $f,g \in L^{2}(\RN)$.
By Riesz-Thorin interpolation, $\mathcal{F}$ can be extended to any $L^{p}(\RN)$ with $1<p<2$, and for $p>2$ it can be defined in the distributional sense (see, e.g. \cites{Hoermander_LinDifOp,Rudin_FktAnalysis}).

% ==========================================
\subsection{Spaces of Bandlimited Functions}

In many applications, and especially in communications, bandlimited signals are the prevailing signal model.
In order to set our discussion in the context of known results, we consider two families of bandlimited functions\index{Function!bandlimited}, namely Paley-Wiener and Bernstein spaces.

% --------------------------------------------------------
\runinhead{Paley-Wiener Spaces}\index{Space!Paley-Wiener $\PW^{p}_{\sigma}$}
For $\sigma > 0$ and $1 \leq p \leq \infty$, the \emph{Paley-Wiener space} $\PW^{p}_{\sigma}$ is the set of all functions $f$ that can be represented as
\begin{equation}
\label{equ:DefPW}
	f(z) = \frac{1}{2\pi} \int^{\sigma}_{-\sigma} \widehat{f}(\omega)\, \E^{\I\omega z}\, \d\omega\;,
	\qquad z\in \CN	
\end{equation}
for some $\widehat{f} \in L^{p}([-\sigma,\sigma])$.
Thus $f$ can be represented as the inverse Fourier transform of a function $\widehat{f}$ in $L^{p}([-\sigma,\sigma])$, and we say that $f$ has \emph{bandwidth}\index{Bandwidth} $\sigma$.
The norm in $\PW^{p}_{\sigma}$ is defined by $\|f\|_{\PW^{p}_{\sigma}} = \|\widehat{f}\|_{L^{p}([-\sigma,\sigma])}$.

The Paley-Wiener spaces are nested. Indeed, H{\"o}lder's inequality implies that $\|f\|_{\PW^{q}_{\sigma}} \leq \|f\|_{\PW^{p}_{\sigma}}$ for all $f \in \PW^{p}_{\sigma}$ and all $1 \leq q \leq p$.
This yields the inclusions $\PW^{p}_{\sigma} \subset \PW^{q}_{\sigma}$ for all $1 \leq q \leq p < \infty$.
In particular, $\PW^{1}_{\sigma}$ is the largest space in the family of Paley-Wiener spaces.
By the properties of the Fourier transform, it follows easily from \eqref{equ:DefPW} that any Paley-Wiener function $f$ is continuous on $\RN$ with $\|f\|_{\infty} \leq \frac{1}{2\pi}\|f\|_{\PW^{1}_{\sigma}}$
and the \emph{Riemann-Lebesgue lemma}\index{Lemma!Riemann-Lebesgue} shows that any Paley-Wiener function vanishes at infinity such that $\PW^{p}_{\sigma} \subset \mathcal{C}_{0}(\RN)$ for every $1 \leq p \leq \infty$.

Similarly, as for the $L^{p}(\RN)$ spaces, the Paley-Wiener space $\PW^{2}_{\sigma}$ plays a particular role since it is a Hilbert space with the $L^{2}$ inner product given on the left-hand side of \eqref{equ:InnerProductL2}.
Moreover, it is even a reproducing kernel Hilbert space.
This means that for every $\lambda\in\RN$, and for all $f \in \PW^{2}_{\sigma}$, the point evaluation $f(\lambda)$ can be written as an inner product 
\begin{equation*}
	f(\lambda) = \left\langle f , r_\lambda\right\rangle
	\qquad\text{with}\qquad
	r_{\lambda}(t) = \frac{\sin(\sigma[t-\lambda])}{\pi[t-\lambda]}\;,
\end{equation*}
and with the \emph{reproducing kernel}\index{Reproducing kernel} $r_{\lambda} \in \PW^{2}_{\sigma}$ with $\|r_{\lambda}\|_{\PW^{2}_{\pi}} = \sqrt{\sigma/\pi}$.

% --------------------------------------------------
\runinhead{Bernstein Spaces}\index{Space!Bernstein $\B^{p}_{\sigma}$}
For any $\sigma > 0$, the set $\B_{\sigma}$ contains all entire functions\index{Function!entire} of exponential type $\leq \sigma$, i.e. to every $f \in \B_{\sigma}$ and every $\epsilon > 0$ there is a constant $C = C(f,\epsilon)$ such that 
\begin{equation*}
	|f(z)| \leq C\, \E^{(\sigma+\epsilon) |z|}
	\quad\text{for all}\quad z\in\CN\;.
\end{equation*}
Then for $1 \leq p \leq \infty$, the \emph{Bernstein space} $\B^{p}_{\sigma}$ is the set of all $f\in\B_{\sigma}$ whose restriction to the real axis belongs to $L^{p}(\RN)$.
The norm in $\B^{p}_{\sigma}$ is defined as the $L^{p}(\RN)$ norm of $f$ on $\RN$.

Functions in the Bernstein spaces $\B^{p}_{\sigma}$ are also bandlimited in the sense that they have a Fourier transform with finite support.
This follows from the \emph{Paley-Wiener theorem}\index{Theorem!of Paley-Wiener} \cites{Hoermander_LinDifOp,Rudin_FktAnalysis}. It states that $f$ is an entire function of exponential type $\leq \sigma$, if and only if it is the Fourier transform of a distribution with compact support which is contained in the interval $[-\sigma,\sigma]$.
Finally, we remark that the \emph{theorem of Plancherel-P{\'o}lya}\index{Theorem!of Plancherel-P{\'o}lya} implies that there exists a constant $C = C(p,\sigma)$, such that for all $f \in \B^{p}_{\sigma}$
\begin{equation}
\label{equ:PlPol}
	|f(t + \I\tau)| \leq C\, \|f\|_{p}\, \E^{\sigma |\tau|}
	\quad\text{for all}\ t,\tau\in\RN\;.
\end{equation}
Thus every $f \in \B^{p}_{\sigma}$ is uniformly bounded on every line parallel to the real axis.
It follows that $\B^{p}_{\sigma} \subset \B^{q}_{\sigma}\subset \B^{\infty}_{\sigma}$ for all $1\leq p \leq q \leq \infty$.
In particular, $\B^{\infty}_{\sigma}$ is the largest space in the family of Bernstein spaces.
The space of all functions $f \in \B^{\infty}_{\sigma}$ for which $f(t) \to 0$ as $t \to \pm \infty$ will be denoted by $\B^{\infty}_{\sigma,0}$, and we have the relations $\B^{\infty}_{\sigma,0} \subset \mathcal{C}_{0}(\RN)$ and $\B^{p}_{\sigma} \subset \mathcal{C}(\RN)$ for every $1 \leq p \leq \infty$.
Note also that Plancherel's theorem shows that $\B^{2}_{\sigma} = \PW^{2}_{\sigma}$. 
So overall we have the relation
\begin{equation*}
	\B^{2}_{\sigma} = \PW^{2}_{\sigma} \subset \PW^{1}_{\sigma} \subset \B^{\infty}_{\sigma}\;.
\end{equation*}
Without any loss of generality, we normalize the bandwidth $\sigma$ of our signals to $\sigma = \pi$ throughout this article.

In applications, it is often important to control the peak value of the signals because of limited dynamic ranges of power amplifiers and other hardware components \cite{Wunder13}.
Moreover, the energy efficiency is an increasingly important aspect for mobile communication networks, and
since the efficiency of high power amplifiers is directly related to the peak-to-average power ratio\index{Peak-to-average power ratio} of the signals, it is necessary to control the peak values of the signals to design energy efficient systems. For such applications, $\B^{\infty}_{\pi}$ is the appropriated signal space.
Moreover, in sampling and reconstruction of stochastic processes, the space $\PW^{1}_{\pi}$ plays a fundamental role \cites{Butzer_DMV88,BoMo_IEEESIT10} because the spectral densities of such processes are $L^1$ functions, in general. Consequently, one has to investigate the behavior of the reconstruction series for functions in $\PW^{1}$.
For these reasons, the spaces $\B^{\infty}_{\pi}$ and $\PW^{1}_{\pi}$ are the primary signal spaces considered in this paper.

% =================================================================================
% ========== Sampling Indroduction ================================================
% =================================================================================
\section{Classical Sampling Theory -- A Short Introduction}
\label{sec:SampTheory}

Sampling theory deals with the reconstruction of functions $f$ in terms of their values (\emph{samples}) $f(\lambda_{n})$ on an appropriated set $\{\lambda_{n}\}$ of sampling points.
The particular choice of the set $\{\lambda_{n}\}$ and in particular the density of the points $\lambda_{n}$ determines whether it is possible to reconstruct $f$ from its samples \cite{Seip}.
This theory is the foundation of all modern digital signal processing \cite{Shannon_IRE1949}. Moreover, in \cite{Feynman_LecturesComp}, \emph{Feynman} discusses sampling theory in a much wider context (\emph{``physics of computations''}) and its importance for theoretical physics.
This section reviews some of the most important results in sampling theory, as far as they will be needed in the subsequent discussions.

% ----- Unoform Sampling Series -----
\subsection{Uniform Sampling}

We start our discussion with the situation where the sampling points $\Lambda = \{\lambda_{n}\}_{n\in\ZN}$ are distributed uniformly on the real axis, i.e. where $\lambda_{n} = n$
 for all $n\in\ZN$.
Then the fundamental initial result in sampling theory is the so-called \emph{cardinal series}\index{Series!cardinal} \cite{Higgens_ShortStories85} which is also known by the name \emph{Whittaker-Shannon-Kotelnikov sampling theorem}.
Let $f \in \PW^{p}_{\pi}$ be a bandlimited function and consider the \emph{Shannon series} of degree~$N$ 
\begin{equation}\index{Series!Shannon}
\label{equ:ShanSampSer}
	(\Op{S}_{N}f)(t)
	= \sum^{N}_{n=-N} f(n) r_{n}(t)
	= \sum^{N}_{n=-N} f(n) \frac{\sin(\pi[t-n])}{\pi[t-n]}
\end{equation}
with the reproducing kernels $r_{n}$ of $\PW^{2}_{\pi}$.
This series is intended to approximate the given function $f$ from its samples $\{f(n)\}^{N}_{n=-N}$.
The question is whether, and in which sense, $\Op{S}_{N}f$ converges to the given function $f$ as $N\to\infty$.

Original research was mainly focused on functions in the Hilbert space $\PW^{2}_{\pi}$, and it is well known that
\begin{equation}
\label{equ:ConVergPW2}
	\lim_{N\to\infty} \left\| f - \Op{S}_{N}f \right\|_{\PW^{2}_{\pi}} = 0
	\quad\text{for all}\quad f \in \PW^{2}_{\pi}\;.
\end{equation}
This result easily follows by observing that the reproducing kernels $\{ r_{n} \}_{n\in\ZN}$ form an orthonormal basis for $\PW^{2}_{\pi}$ \cite{Hardy_41}. Moreover, since $\PW^{2}_{\pi}$ is a reproducing kernel Hilbert space, Cauchy-Schwarz inequality immediately gives
\begin{equation*}
	\left|f(t) - (\Op{S}_{N}f)(t) \right|
	= \left| \left\langle f - \Op{S}_{N}f , r_t \right\rangle \right|
	\leq \|r_t\|_{\PW^{2}_{\pi}}\, \| f - \Op{S}_{N}f \|_{\PW^{2}_{\pi}}\;.
\end{equation*}
Since $\|r_t\|_{\PW^{2}_{\pi}} = 1$ for all $t\in\RN$, \eqref{equ:ConVergPW2} also implies
\begin{equation*}
	\lim_{N\to\infty} \max_{t\in\RN} \left| f(t) - (\Op{S}_{N}f)(t) \right| = 0
	\quad\text{for all}\quad f \in \PW^{2}_{\pi}\;.
\end{equation*}
In other words, $\Op{S}_{N} f$ converges uniformly to $f$ for all $f \in \PW^{2}_{\pi}$, and one can even show that $S_{N}f$ converges absolutely.
Moreover, it is fairly easy to extend the above results for $\PW^{2}_{\pi}$ to all Paley-Wiener space $\PW^{p}_{\pi}$ with $1 < p \leq \infty$.
More precisely, one has the following statement \cite{Higgins_Sampling1}:

\begin{theorem}
\label{thm:standartPW}
For each $1 < p \leq \infty$ and for all $f \in \PW^{p}_{\pi}$, we have 
\begin{equation*}
	f(t)
	= \lim_{N\to\infty} (\Op{S}_{N}f)(t)
	= \sum^{\infty}_{n=-\infty} f(n) \frac{\sin(\pi[t-n])}{\pi[t-n]}
\end{equation*}
where the sum converges absolutely and uniformly on the whole real axis $\RN$,
and also in the norm of $\PW^{p}_{\pi}$.
\end{theorem}

Theorem~\ref{thm:standartPW} gives a simple and convenient reconstruction formula for all functions in the Paley-Wiener spaces $\PW^{p}_{\pi}$ with $p>1$.
However, we want to stress that Theorem~\ref{thm:standartPW} does not hold for the largest Paley-Wiener space $\PW^{1}_{\pi}$.
The convergence of the sampling series in this space will be considered in more detail in Section~\ref{sec:GUConv}. In particular, Corollary~\ref{cor:ConvPW1oversamp} below will show that with oversampling, $\Op{S}_{N}$ converges uniformly on $\PW^{1}_{\pi}$.

% ---------------------------------------
% ----- Non-Uniform Sampling Series -----
% ---------------------------------------
\subsection{Non-Uniform Sampling Series}

The Shannon sampling series \eqref{equ:ShanSampSer} is based on uniform signal samples taken at integer values $\lambda_{n} = n$, $n\in\ZN$.
To gain more flexibility, one may consider series which reconstruct a function $f$ from samples $\{ f(\lambda_{n}) \}_{n\in\ZN}$ taken at a set $\Lambda = \{\lambda_{n}\}_{n\in\ZN}$ of non-uniform sampling points.
To choose an appropriate sampling set $\Lambda$, we start again with the Hilbert space $\B^{2}_{\pi} = \PW^{2}_{\pi}$. 
For this signal space, so-called complete interpolating sequences are suitable sampling sets.

\begin{definition}[Interpolating and Sampling Sequence]
Let $\Lambda = \{\lambda_n\}_{n\in\ZN}$ be a sequence in $\CN$ and let $\Op{S_{\Lambda}} : \B^{2}_{\pi} \to \ell^{2}$ be the associated sampling operator defined by
$\Op{S_{\Lambda}} : f \mapsto \{f(\lambda_{n})\}_{n\in\ZN}$.
We call $\Lambda$
\begin{itemize}
\item a \emph{sampling sequence}\index{Sequence!sampling} for $\B^{2}_{\pi}$ if $\Op{S_{\Lambda}}$ is injective.
\item an \emph{interpolation sequence}\index{Sequence!interpolating} for $\B^{2}_{\pi}$ if $\Op{S_{\Lambda}}$ is surjective.
\item \emph{complete interpolating}\index{Complete interpolating} for $\B^{2}_{\pi}$ if $\Op{S_{\Lambda}}$ is bijective.
\end{itemize}
\end{definition}

In the following, we always use as sampling sets complete interpolating sequences for $\B^{2}_{\pi}$.
Such sequences were completely characterized by \emph{Minkin} \cite{Minkin_92} after \cites{Pavlov_79,Nikolskii_80} already gave characterizations under mild constrains on $\Lambda$.

\runinhead{Over- and undersampling}
Assuming that $\Lambda$ is complete interpolating for $\B^{2}_{\pi}$, then $\Op{S_\Lambda}$ is an isomorphism between the signal space $\B^{2}_{\pi}$ and the sequence space $\ell^{2}$ such that the interpolation problem $f(\lambda_n) = c_{n}$, $n\in\ZN$ has a unique solution $f \in \B^{2}_{\pi}$ for every sequence $\{c_n\}_{n\in\ZN} \in \ell^{2}$.
Now, let $\beta \in \RN$ and consider the signal space $\B^{2}_{\beta \pi}$.
If $\beta > 1$, then $\B^{2}_{\pi}$ is a proper subset of $\B^{2}_{\beta\pi}$ and $\Op{S_{\Lambda}}$ will no longer be injective viewed as an operator on $\B^{2}_{\beta\pi} \to \ell^{2}$.
Therefore it will not be possible to reconstruct every $f \in \B^{2}_{\beta\pi}$ uniquely from its samples $\Op{S_{\Lambda}}f$. In this case, we say that $\B^{2}_{\beta\pi}$ is \emph{undersampled} by $\Lambda$.
Conversely, if $\beta < 1$, then $\B^{2}_{\beta\pi}$ is a proper subset of $\B^{2}_{\pi}$ and the sampling operator $\Op{S_{\Lambda}} : \B^{2}_{\beta\pi} \to \ell^{2}$ is injective but not surjective.
In this case, every function $f \in \B^{2}_{\beta\pi}$ can uniquely be reconstructed from its samples $\Op{S_{\Lambda}}$ but there exist sequences $\{c_{n}\}_{n\in\ZN} \in \ell^{2}$ such that the interpolation problem $f(\lambda_n) = c_{n}$, $n\in\ZN$ has no solution in $\B^{2}_{\beta\pi}$.
In this case, we say that $\B^{2}_{\beta\pi}$ is \emph{oversampled} by $\Lambda$ and $1/\beta$ is the \emph{oversampling factor}.

Let $\Lambda = \{\lambda_n\}_{n\in\ZN}$ be a complete interpolating sequence for $\B^{2}_{\pi}$ and define the function
\begin{equation}
\label{equ:DefVarPhi}
   \varphi(z) = z^{\delta_{\Lambda}}\lim_{R\to\infty} \prod_{\substack{|\lambda_{n}|<R\\\lambda_{n}\neq 0}} \left( 1-\frac{z}{\lambda_{n}} \right)
   \qquad\text{with}\qquad
   \delta_{\Lambda}
   = \left\{\begin{array}{ll}
     1\ &\ \text{if}\ 0 \in \Lambda\\
	0\ &\ \text{otherwise}
   \end{array}\right.\;.
\end{equation}
One can show \cites{Levin1997_Lectures,Young_NonHarmonic} that the product in \eqref{equ:DefVarPhi} converges uniformly on every compact subset of $\CN$ and that $\varphi$ is an entire function of exponential type $\pi$.
It follows from \eqref{equ:DefVarPhi} that $\varphi(\lambda_{n}) = 0$, i.e. $\Lambda$ is the zero set of the function $\varphi$
which is often called the \emph{generating function}\index{Function!generating} of $\Lambda$.
Based on the function \eqref{equ:DefVarPhi}, one defines for every $n\in\ZN$ %the functions
\begin{equation}
\label{equ:Kernels}
	\varphi_{n}(z) := \frac{\varphi(z)}{\varphi'(\lambda_{n}) (z-\lambda_{n})}\;,
	\qquad z\in\CN\;.
\end{equation}
Again, these are entire functions of exponential type $\pi$ which solve the interpolation problem $\varphi_{n}(\lambda_{n}) = 1$ and $\varphi_{n}(\lambda_{k}) = 0$ if $k \neq n$.
Following the ideas of classical Lagrange interpolation, one considers for any $N\in\NN$, the approximation operator
\begin{equation}
\label{equ:NonUnifSS}
	(\Op{A}_{N}f)(z) = \sum^{N}_{n=-N} f(\lambda_{n})\, \varphi_{n}(z)\;,
\end{equation}
which only involves $2N+1$ sampling values.
The aim is to approximate any function $f \in \B^{2}_{\pi}$ by $\Op{A}_{N} f$ in such a way that the approximation error $\|f - \Op{A}_{N} f\|_{\B^{2}_{pi}}$ becomes less than any arbitrary given bound $\epsilon$ as soon as $N\in\NN$ is sufficiently large.

By the definition of the interpolation kernels $\varphi_{n}$, it is clear that $\Op{A}_{N}f$ satisfies the interpolation condition $(\Op{A}_{N}f)(\lambda_{n}) = f(\lambda_{n})$ for all $n = 0,\pm 1,\pm 2,\dots,\pm N$, and one can show that for every $f \in \B^{2}_{\pi}$, the sequence $\{ \Op{A}_{N}f \}_{N\in\NN}$ converges in $\B^{2}_{\pi}$ as $N\to\infty$.
Since $\Lambda$ is completely interpolating, we therefore have
$\Op{A}_{N} f$ converging to $f$ for every $f \in \B^{2}_{\pi}$ \cite{Young_NonHarmonic}.

% ----- Thm: Non-Equidistant PW2 -----
\begin{theorem}
\label{thm:NonUniformSamp}
Let $\Lambda = \{\lambda_{n}\}_{n\in\ZN}$ be a complete interpolating sequence for $\B^{2}_{\pi}$ and let $\{\varphi_{n}\}_{n\in\ZN}$ be the functions defined by \eqref{equ:Kernels} based on the generating function \eqref{equ:DefVarPhi}. Then 
\begin{equation*}
	f(z)
	= \lim_{N\to\infty} (\Op{A}_{N} f)(z)
	= \sum^{\infty}_{n=-\infty} f(\lambda_{n})\, \varphi_{n}(z)
\end{equation*}
for all $f \in \B^{2}_{\pi}$ where the sum converges in the norm of $\B^{2}_{\pi}$, and uniformly on $\RN$.
\end{theorem}
% ------------------------------------

In general, the characterization of complete interpolating sequences $\Lambda$ is fairly complicated and the calculation of the corresponding generating function $\varphi$ via \eqref{equ:DefVarPhi} can be computationally difficult.
Fortunately, an important subset of complete interpolating sequences is known which simplifies the entire procedure considerably.
These are the zero sets of the so-called sine-type functions.

\begin{definition}[Sine-type function]\index{Function!sine-type}
An entire function $\varphi$ of exponential type $\pi$ is said to be a \emph{sine-type function} if it has simple and separated zeros and if there exist positive constants $A,B,H$ such that
\begin{equation*}
	A\, \E^{\sigma |\eta|} 
	\leq |\varphi(\xi + \I \eta)|
	\leq B\, \E^{\sigma |\eta|}
	\quad\text{for all}\quad
	\xi\in\RN\ \text{and}\ |\eta| \geq H\;.
\end{equation*}
\end{definition} 
Any sine-type function can be determined from its zero set $\Lambda$ by \eqref{equ:DefVarPhi}.

\begin{example}
The uniform sampling considered above is a special case of non-uniform sampling.
The sampling set $\Lambda$ is obtained as the zero set of the sine-type function $\varphi(z) = \sin(\pi z)$, which is equal to $\Lambda = \{\lambda_{n} = n\}_{n\in\ZN}$. The corresponding interpolation kernels \eqref{equ:Kernels} become $\varphi_{n}(z) = \sin(\pi[z-n])/(\pi[z-n])$ and \eqref{equ:NonUnifSS} becomes equal to \eqref{equ:ShanSampSer}.
\end{example}

If the sampling set $\Lambda$ is chosen as the zero set of a sine-type function, then Theorem~\ref{thm:NonUniformSamp} can be extended to all Bernstein spaces $\B^{p}_{\pi}$ with $1 < p < \infty$.
Thus, the non-uniform sampling series \eqref{equ:NonUnifSS} reconstructs every function in $\B^{p}_{\pi}$. More precisely, one has the following statement \cite{Levin1997_Lectures}*{Lect.~22}.

% ----- THEOREM: non-uniform Sampling Bernstein -----
\begin{theorem}
\label{thm:StandartBp}
Let $\Lambda = \{\lambda_{n}\}_{n\in\ZN}$ be the zero set of a sine-type function $\varphi$ and let $\{\varphi_{n}\}_{n\in\ZN}$ and $\Op{A}_{N}$ be defined as in \eqref{equ:Kernels} and \eqref{equ:NonUnifSS}, respectively.
Then for each $1 \leq p < \infty$
\begin{equation*}
	f(t)
	= \lim_{N\to\infty} (\Op{A}_{N}f)(t)
	= \sum^{\infty}_{n=-\infty} f(\lambda_{n})\, \varphi_{n}(t)\;,
	\qquad\text{for all}\ f \in \B^{p}_{\pi}
\end{equation*}
where the sum converges uniformly on $\RN$ and for $1 < p < \infty$, it also converges in the norm of $\B^{p}_{\pi}$.
\end{theorem}
% ---------------------------------------------------

Also here we would like to stress that Theorem~\ref{thm:StandartBp} does not hold for the largest space $\B^{\infty}_{\pi}$ in the family of Bernstein spaces, and we will discuss the $\B^{\infty}_{\pi}$-case later in Sec.~\ref{sec:GUConv} (cf. Conjecture~\ref{con:Conj1} and Theorem~\ref{thm:ConvergBinfty} below).

% ==============================================
% ========== Uniform Convergence ===============
% ==============================================
\section{On the Global Uniform Convergence of Sampling Series}
\label{sec:GUConv}

Theorems~\ref{thm:standartPW} and \ref{thm:StandartBp} establish the uniform convergence of the sampling series on $\PW^{p}_{\pi}$ for $1 < p \leq \infty$ and on $\B^{p}_{\pi}$  for $1 \leq p < \infty$, respectively.
However, both results can not easily be extended to the largest spaces $\PW^{1}_{\pi}$ and $\B^{\infty}_{\pi}$.
This section reviews and discusses some recent results which investigate on which signal spaces and under which conditions the sampling series converges uniformly on $\RN$.
So we are going to investigate the behavior of the quantity 
\begin{equation*}
	\max_{t\in\RN} \left|f(t) - (\Op{S}_{N}f)(t) \right|
	\qquad\text{or}\qquad
	\max_{t\in\RN} \left|f(t) - (\Op{A}_{N}f)(t) \right|	
\end{equation*}
as $N$ tends to infinity.
This quantity is an important measure for the stability of the reconstruction process since it allows us to control the peak value of the error between the approximation $\Op{A}_{N}f$ and the function $f$ itself.
The question is whether the maximum error can be made arbitrarily small for a sufficiently large approximation degree $N$.

For the signal space $\PW^{1}_{\pi}$, a classical theorem due to \emph{Brown} states that the Shannon sampling series $\Op{S}_{N}$ converges uniformly on compact sets of $\RN$.

% ----- Thm: Brown -----
\begin{theorem}[Brown \cite{Brown_67}]
\label{thm:Brown}
For all $f \in \PW^{1}_{\pi}$ and for all $T>0$, we have
\begin{equation*}
	\lim_{N\to\infty} \left( \max_{t \in [-T,T]} \left| f(t) - (\Op{S}_{N}f)(t) \right|\right) = 0\;.
\end{equation*}
\end{theorem}
Based on this result we consider now the following two questions.
\begin{enumerate}
\item
Is it possible to have even uniform convergence on the entire real axis, i.e. is it possible to replace the interval $[-T,T]$ by $\RN$ in Theorem~\ref{thm:Brown}?
\item 
Is it possible to extend Theorem~\ref{thm:Brown} to the larger space $\B^{\infty}_{\pi}$ of bounded bandlimited functions?
\end{enumerate}
Since Theorem~\ref{thm:Brown} is based on uniform sampling without oversampling, we may hope to achieve these extensions 
by replacing the uniform sampling series $\Op{S}_{N}$ with a non-uniform series $\Op{A}_{N}$ and by using oversampling.

% ----- uniform Sampling ------
\subsection{Weak Divergence of the Shannon Sampling Series}
\label{sec:Divergence}

We begin by asking whether the Shannon sampling series \eqref{equ:ShanSampSer} converges uniformly on the whole real axis $\RN$ for every function $f \in \PW^{1}_{\pi}$.
The negative answer is given by the following theorem \cite{BoMo_IEEESP08}.

% ----- Thm: Weak Divergence -----
\begin{theorem}
\label{thm:GlobalDivShannon}
There exists a signal $f_{0} \in \PW^{1}_{\pi}$ such that
\begin{equation}
\label{equ:LimSupDiv}
	\limsup_{N\to\infty}\ \| \Op{S}_{N} f_{0} \|_{\infty} = \infty\;.
\end{equation}
\end{theorem}
% --------------------------------

\begin{remark}
Since $\PW^{1}_{\pi} \subset \mathcal{C}_{0}(\RN)$, all functions in $\PW^{1}_{\pi}$ are bounded on $\RN$. Therefore Theorem~\ref{thm:GlobalDivShannon} implies in particular that there exists an $f_{0} \in \PW^{1}_{\pi}$ such that
$\limsup_{N\to\infty} \|f_{0} - \Op{S}_{N} f_{0}\|_{\infty} = \infty$.
\end{remark}

\begin{remark}
In fact, the divergence behavior described by Theorem~\ref{thm:GlobalDivShannon} is not a particular property of the Shannon sampling series but holds for a large class of approximation processes which rely on uniform sampling.
More precisely, \cite{BoMo_IEEESP08} proved Theorem~\ref{thm:GlobalDivShannon} not only for the sampling series \eqref{equ:ShanSampSer} but for all sampling series with the general form
\begin{equation}
\label{equ:SampSerR}
	(\Op{R}_{N}f)(t)
	= (\Op{T} f)(t) + \sum^{N}_{n=-N} f(n)\, \phi_{n}(t)\;,
\end{equation}
where $\Op{T} : \PW^{1}_{\pi} \to \B^{\infty}_{\pi}$ is linear and bounded, and $\phi_{n} \in \B^{\infty}_{\pi}$ are certain interpolation kernels.
If the series $\Op{R}_{N}$ satisfies\footnote{Interpolation series $\Op{R}_{N}$ which satisfy these conditions include the so called \emph{Valiron series}\index{Series!Valiron} \cite{Boas_EntireFunct54,Higgens_ShortStories85,Valiron_Interpol25} or \emph{Tschakaloff series}\index{Series!Tschakaloff} \cite{Higgens_ShortStories85,Tschakaloff33}.
} the following three properties:
\begin{itemize}
\item
The kernels $\phi_{n}$ are uniformly bounded, i.e.  $\|\phi_{n}\|_{\infty} \leq C_{\phi} <\infty$ for all $n\in\ZN$,
\item
$(\Op{R}_{N}f)(t)$ converges pointwise to $f(t)$ for all $f \in \PW^{2}_{\pi}$,
\item
The operator $\Op{T}$ is such that there exist two constants $C, D > 0$ such that for all $f\in \PW^{1}_{\pi}$ always $\sup_{t\in\RN} |(\Op{T}f)(t)| \leq C\, \max_{|z|\leq D}|f(z)|$,
\end{itemize}
then it shows the same divergence behavior as in Theorem~\ref{thm:GlobalDivShannon}. 
Moreover, the particular function $f_{0} \in \PW^{1}_{\pi}$, for which $\|\Op{S}_{N}f_{0}\|_{\infty}$ diverges, is universal in the sense that all interpolation series $\Op{R}_{N}$ with the above properties diverge for $f_{0}$.
\end{remark}

The proof of Theorem~\ref{thm:GlobalDivShannon} relies on an explicit construction of $f_{0} \in \PW^{1}_{\pi}$ and a corresponding subsequence $\{N_{k}\}^{\infty}_{k=1}$ such that
\begin{equation*}
	(\Op{S}_{N_{k}} f_{0})(N_{k} + 1/2) \geq C_{1}\, \sqrt{k^{3}} + C_{2} \to \infty
	\quad\text{as}\ k\to\infty\;.
\end{equation*} 
Because of this construction, one has the $\limsup$-divergence in \eqref{equ:LimSupDiv}.
In a sense, this is a weak notion of divergence, because one designs a very specific function $f_{0}$ and a corresponding subsequence of approximations $\{\Op{S}_{N_k} f_{0}\}_{k\in\NN}$ such that divergence emerges.
This notion of (weak) divergence is sufficient to show that the approximation procedure is not always convergent.
However, it does not show that there exists no recovery procedure at all.
In particular, the divergence result of Theorem~\ref{thm:GlobalDivShannon} does not allow to answer the following two questions:

\begin{description}[Type 1]
\item[Q-1]
Let $f \in \PW^{1}_{\pi}$ be arbitrary. Does there exist a specific sequence $\mathcal{N}(f) = \{N_{k} = N_{k}(f)\}_{k\in\NN}$, depending on $f$, such that
\begin{equation}
\label{equ:ConvergendSubSequ}
	\sup_{k\in\NN} \|f - \Op{S}_{N_k} f\|_{\infty} < \infty\;.
\end{equation}

\item[Q-2]
Does there exist a universal approximation sequence $\mathcal{N} = \{N_{k}\}_{k\in\NN}$ such that \eqref{equ:ConvergendSubSequ} holds for all $f \in \PW^{1}_{\pi}$?
\end{description}

\begin{remark}
Note that a negative answer to Q-1 implies a negative answer to Q-2. Conversely, a positive answer to Q-2 implies a positive answer to Q-1.
\end{remark}

With that said, Theorem~\ref{thm:GlobalDivShannon} gives only a weak statement about the global divergence behavior of the Shannon series on $\PW^{1}_{\pi}$.
Because, if Q-2 were to have a positive answer, then one would have a globally convergent method to reconstruct every $f \in \PW^{1}_{\pi}$ from its sampled values $\{f(n)\}_{n\in\ZN}$.
But even if only question Q-1 were to have a positive answer, signal recovery would still be possible, but with an adaptive approximation process which depends on the actual function.

Divergence results like those in Theorem~\ref{thm:GlobalDivShannon} are usually proved using the \emph{uniform boundedness principle}\index{Uniform boundedness principle}.
This principle is one of the cornerstones of functional analysis and it may be formulated as follows (see, e.g., \cite{Rudin}):

% ----- Theorem Banach-Steinhaus -----
\begin{theorem}[Banach-Steinhaus, \cite{BanachSteinhaus27}]\index{Theorem!of Banach-Steinhaus}
\label{thm:BS}
Let $\{ \Op{T}_{n} \}_{n\in\NN}$ be a sequence of linear operators $\Op{T}_{n} : \mathcal{X} \to \mathcal{Y}$ mapping a Banach space $\mathcal{X}$ into a normed space $\mathcal{Y}$ with the operator norms
\begin{equation*}
	\| \Op{T}_{n} \| = \sup_{f\in\mathcal{X}} \frac{ \|\Op{T}_{n} f\|_{\mathcal{Y}} }{ \|f\|_{\mathcal{X}} }\;.
\end{equation*}
If $\sup_{n\in\NN}  \|\Op{T}_{n}\| = \infty$
%\begin{equation*}
	%\sup_{n\in\NN}  \|\Op{T}_{n}\| = \infty
%\end{equation*}
then there exists an $x_{0} \in \mathcal{X}$ such that
\begin{equation}
\label{equ:BSTheorem}
	\sup_{n\in\NN} \|\Op{T}_{n} x_{0}\|_{\mathcal{Y}} = \infty\;.
\end{equation}
In fact, the set $\mathcal{D}$ of all $x_{0} \in \mathcal{X}$ which satisfy \eqref{equ:BSTheorem} is a residual set $\mathcal{X}$.
\end{theorem}
% ------------------------------------

\begin{remark}
In a Banach space, a \emph{residual set}\index{Residual set} is the complement of a set of first category (a meager set) and therefore it is a set of second category (i.e. a nonmeager set).
In the following we use that the countable intersection of residual sets is again a residual set. In particular, the countable intersection of open dense subsets is a residual set \cite{Kantorovich}.
\end{remark}

Here, we shortly discuss how the theorem of Banach-Steinhaus can be used to investigate the two questions Q-1 and Q-2.
In particular, we want to show the limitations of the Banach-Steinhaus theorem for answering question Q-1.

To prove Theorem~\ref{thm:GlobalDivShannon}, based on the uniform boundedness principle, it is sufficient to shown that the norms
\begin{equation*}
	\|\Op{S}_{N}\|
	= \sup\left\{ \|\Op{S}_{N} f\|_{\infty}\ :\  f\in\PW^{1}_{\pi}\ ,\ \|f\|_{\PW^{1}_{\pi}} \leq 1 \right\}
\end{equation*}
of the operators $\Op{S}_{N} : \PW^{1}_{\pi} \to \B^{\infty}_{\pi}$, defined in \eqref{equ:ShanSampSer}, are not uniformly bounded. 
This was done in \cite{BoMo_IEEESP08}, where it was shown that there exists a constant $C_{S}$ such that
\begin{equation}
\label{equ:LebesgueKonstantSN}
	\|\Op{S}_{N}\| \geq C_{S}\, \log N
	\qquad\text{for all}\ N\in\NN\;.
\end{equation}
Then Theorem~\ref{thm:BS} implies immediately that there exits a residual set $\mathcal{D} \subset \PW^{1}_{\pi}$ such that
\begin{equation*}
	\limsup_{N\to\infty} \| \Op{S}_{N} f \|_{\infty} = + \infty
	\quad\text{for all}\quad
	f \in \mathcal{D}\;.
\end{equation*}

Next we use Theorem~\ref{thm:BS} to investigate question Q-2.
Since \eqref{equ:LebesgueKonstantSN} holds for all $N \in \NN$, the same reasoning can be applied to any subsequence $\mathcal{N} = \{N_{k}\}_{k\in\NN}$ of $\NN$.
Then the Banach-Steinhaus theorem states that there exists a residual set $\mathcal{D}(\mathcal{N}) \subset \PW^{1}_{\pi}$ such that
\begin{equation*}
	\limsup_{k\to\infty} \|\Op{S}_{N_{k}} f\|_{\infty} = +\infty
	\quad\text{for all}\quad
	f \in \mathcal{D}(\mathcal{N})\;.
\end{equation*}
This shows that the answer to question Q-2 is actually negative, i.e. there exists no universal subsequence $\mathcal{N} = \{N_{k}\}_{k\in\NN}$ such that $\Op{S}_{N_k}f$ converges uniformly for every $f \in \PW^{1}_{\pi}$.
One can even say more about the size of the divergence set.
Let $\{\mathcal{N}_{v}\}_{v\in\NN}$ be a countable collection of subsequences of $\NN$.
Then to every $\mathcal{N}_{v} = \{N_{v,k}\}_{k\in\NN}$ there exists a subset $\mathcal{D}(\mathcal{N}_{v}) \subset \PW^{1}_{\pi}$ such that
\begin{equation*}
	\limsup_{k\to\infty} \|\Op{S}_{N_{v,k}} f\|_{\infty} = +\infty
	\quad\text{for all}\quad
	f \in \mathcal{D}(\mathcal{N}_{v})\;.
\end{equation*}
Since each $\mathcal{D}(\mathcal{N}_{v})$ is a residual sets in $\PW^{1}_{\pi}$ and since we have only countably many sets, \emph{Baire's category theorem}\index{Theorem!Baire's category} (see, e.g., \cite{Rudin}) implies that the intersection of these sets
\begin{equation}
\label{equ:CountableUnion}
	\bigcap_{v} \mathcal{D}(\mathcal{N}_{v}) \neq \varnothing
\end{equation}
is again a (non-empty) residual set in $\PW^{1}_{\pi}$.
So given a countable collection of subsets $\{\mathcal{N}_v\}_{v\in\NN}$, the set of functions $f \in \PW^{1}_{\pi}$ for which
\begin{equation*}
	\limsup_{k\to\infty} \|\Op{S}_{N_{v,k}} f \|_{\infty} = +\infty
	\quad\text{for all}\quad
	\mathcal{N}_{v} = \{N_{v,k}\}_{k\in\NN} \in \{\mathcal{N}_v\}_{v\in\NN}
\end{equation*}
is nonmeager (of second category) in $\PW^{1}_{\pi}$.

However, the above reasoning cannot be extended to give a definite answer to question Q-1.
Because for a negative answer to Q-1, we need to show that 
\begin{equation*}
	\bigcap_{\substack{\mathcal{N}_v = \{N_{v,k}\}_{k\in\ZN}\\\mathcal{N}_v\ \text{is a subsequence of $\NN$}}} \!\!\!\!\!\!\!\!\! \mathcal{D}(\mathcal{N}_{v})
	\neq \varnothing\;.
\end{equation*}
In other words, we have to show that there exists a function $f_{*} \in \PW^{1}_{\pi}$ such that $\lim_{k\to\infty} \|\Op{S}_{N_{v,k}} f_{*}\|_{\infty} = + \infty$ for \emph{every} subsequence $\mathcal{N}_{v} = \{N_{v,k}\}_{k\in\NN}$ of $\NN$.
However, in contrast to \eqref{equ:CountableUnion}, the set of all subsequences of $\NN$ contains uncountably many elements and the uncountable intersection of residual sets may not be of second category. It even may be empty. Therefore, using the above technique, it is not possible to decide whether this intersection is empty or not.
This way we are not able to answer question Q-1.

In the next subsection we are going to investigate question Q-1 for the Shannon sampling series in more detail, using completely new techniques.
Before that, we give an example of an operator sequence which is (weakly) divergent, but for which question Q-2 can be answered positively.

\begin{example}[Approximation by Walsh functions]
Consider the usual Lebesgue space $L^{2}([0,1])$ of square integrable functions on the interval $[0,1]$
and let $\{\psi_{n}\}^{\infty}_{n=0}$ be the orthonormal system of Walsh functions\index{Function!Walsh} \cite{Walsh_1923} in $L^{2}([0,1])$, where the functions are indexed as in \cite{Fine_WalshFkt_AMS49}.
Let $\Op{P}_{N} : L^{2}([0,1]) \to \cls\{\psi_{n} : n=0,1,\dots,N\}$ be the orthogonal projection onto the first $N+1$ Walsh functions.
Now we view $\Op{P}_{N}$ as a mapping $L^{\infty}([0,1]) \to L^{\infty}([0,1])$ with the corresponding norm
\begin{equation*}
	\|\Op{P}_{N}\| = \sup\left\{ \|\Op{P}_{N}f\|_{\infty}\ :\ f\in L^{\infty}([0,1])\;,\ \|f\|_{\infty} \leq 1 \right\}\;.
\end{equation*}
Then one can show \cite{Fine_WalshFkt_AMS49} that 
\begin{equation*}
	\limsup_{N\to\infty} \|\Op{P}_{N}\| = +\infty
	\qquad\text{but}\qquad
	\|\Op{P}_{2^{k}}\| = 1
	\ \text{for all}\ k\in\NN\;.
\end{equation*}
So the sequence  $\{ \Op{P}_{N} \}_{N\in\NN}$ of linear operators is not uniformly bounded.
Therefore the uniform boundedness principle yields a divergence result similar to Theorem~\ref{thm:GlobalDivShannon}.
However, since there exists a uniformly bounded subsequence $\{ \Op{P}_{2^{k}} \}_{k\in\NN}$, the question Q-2 has a positive answer for this operator sequence $\{P_{N}\}_{N\in\NN}$.
\end{example}

% ===== Strong Divergenz =====
\subsection{Strong Divergence of the Shannon sampling series}
\label{sec:StrongDiv}

The difficulties in answering Q-1, based on the Banach-Steinhaus theorem, may also be viewed as follows.
Under Q-1, the approximation series $\{ N_{k}(f) \}_{k\in\NN}$ can be chosen subject to the actual function $f$, i.e. one is allowed to adapt the reconstruction method to the actual function.
Therefore, the overall approximation procedure $\{\Op{S}_{N_{k}(f)} f\}_{k\in\ZN}$ depends non-linearly on the function $f$.
Hence, one essential requirement for applying the Banach-Steinhaus theorem (the linearity of the operators) is no longer satisfied.
So even though the theorem of Banach-Steinhaus is a very powerful tool for proving (weak) divergence results as in Theorem~\ref{thm:GlobalDivShannon},
it cannot be used to answer question Q-1.
Thus, for the investigation of adaptive recovery algorithms completely different techniques are needed.

We will show below, that for the Shannon sampling series $\Op{S}_{N}$, question Q-1 has a negative answer.
To this end, it is necessary and sufficient to show that the sequence $\{\Op{S}_{N}\}_{N\in\NN}$ diverges \emph{strongly}.

\begin{definition}[Strong divergence]\index{Strong divergence}
Let $\mathcal{X}$ and $\mathcal{Y}$ be Banach spaces, and
let $\{ \Op{T}_{N} \}_{N\in\NN}$ be a sequence of bounded operators $\Op{T}_{N} : \mathcal{X} \to \mathcal{Y}$.
We say that $\Op{T}_{N}$ \emph{diverges strongly} if
\begin{equation*}
	\lim_{N\to\infty} \|\Op{T}_{N} f_{1}\|_{\mathcal{Y}} = \infty
	\quad\text{for some}\quad f_{1}\in\mathcal{X}\;.
\end{equation*}
\end{definition}

So the strong divergence is in contrast to the weaker statement of the $\limsup$ divergence used in Theorem~\ref{thm:GlobalDivShannon}.
As explained above, it is not possible to show the strong divergence of $\Op{S}_{N}$ using the Banach-Steinhaus theorem, and even though several extensions \cite{Stein_AnnMath61,Dickmais1_85,Dickmais2_85} of the Banach-Steinhaus theorem were developed in the past, there currently exists no systematic way to show strong divergence.
For the Shannon sampling series \eqref{equ:ShanSampSer} on $\PW^{1}_{\pi}$, its strong divergence is established by the following theorem.

% ----- Thm: Strong Divergence -----
\begin{theorem}
\label{thm:StrongDiv}
The Shannon sampling series $\Op{S}_{N} : \PW^{1}_{\pi} \to \B^{\infty}_{\pi}$ given in \eqref{equ:ShanSampSer} diverges strongly, i.e.
there exists a function $f_{1} \in \PW^{1}_{\pi}$ for which 
\begin{equation}
\label{equ:StrongDiv}
	\lim_{N\to\infty} \Big(\max_{t\in\RN} \left| (\Op{S}_{N}f_{1})(t) \right|\Big)
	= \lim_{N\to\infty} \|\Op{S}_{N} f_{1}\|_{\infty}
	= \infty\;.
\end{equation}
\end{theorem}
% ----------------------------------

Clearly, this is a much stronger statement than Theorem~\ref{thm:GlobalDivShannon}, with important practical implications for adaptive signal processing algorithms.
It rules out the possibility that the divergence in \eqref{equ:LimSupDiv} occurs only because of a divergent subsequence.
In particular, it implies a negative answer to question Q-1.
Consequently, there exists no (adaptive) signal recovery procedure which converges uniformly on the entire real axis $\RN$.

% === Structure of the divergence set ====
\runinhead{The structure of the divergence sets.}
For non-adaptive linear methods, the Banach-Steinhaus theorem is a very powerful and well established tool in functional analysis to investigate non-adaptive approximation methods. In particular, if one can show that there exists one function $f \in \mathcal{X}$ such that the the sequence $\Op{T}_{N} f$ diverges (weakly) in $\mathcal{Y}$, then the Banach-Steinhaus theorem immediately implies that there exists a whole set $\mathcal{D}_{\mathrm{weak}}$ of second category for which $\T_{N}f$ diverges weakly for every $f \in \mathcal{D}_{\mathrm{weak}}$.

We established in Theorem~\ref{thm:StrongDiv} that there exists one function $f_{1}$ such that the Shannon sampling series diverges strongly.
The question is now whether it is possible to say something about the size or structure of the set of all functions for which $\Op{S}_{N}$ diverges strongly.
Since the Banach-Steinhaus theorem cannot be applied in the case of strong divergence, other techniques have to be developed.
Because of the close relation between strong divergence and adaptive signal processing methods, 
we believe that it is an important research topic to develop general tools for the investigation of strong divergence, similar to the Banach-Steinhaus technique for weak divergence, i.e. for non-adaptive systems.

Here we start with such an investigation and study the structure of the weak and strong divergence sets of approximation series.
To this end, we consider linear approximation operators $\Op{T}_{N} : \mathcal{X} \to \mathcal{Y}$ mapping a Banach space $\mathcal{X}$ into a Banach space $\mathcal{Y}$. Since $\Op{T}_{N} f$ should be a good approximation of $f \in \mathcal{X}$, measured in the topology of $\mathcal{Y}$, it is natural to assume that $\mathcal{X} \subset \mathcal{Y}$.
Additionally, we assume that there exists a dense subset $\mathcal{X}_{0} \subset \mathcal{X}$ such that
\begin{equation}
\label{equ:ConvOnDenseSubSet}
	\lim_{N \to \infty} \|\Op{T}_{N} f_{0} - f_{0}\|_{\mathcal{Y}} = 0
	\quad\text{for all}\ f_{0} \in \mathcal{X}_{0}\;,
\end{equation} 
i.e. such that $\Op{T}_{N} f_{0}$ converges to $f_{0}$ in the norm of $\mathcal{Y}$.
For such linear operators, the next theorem studies the structure of the divergence sets
\begin{equation}
\label{equ:DivSets}
\begin{split}
	\mathcal{D}_{\mathrm{weak}} &= \Big\{ f \in \mathcal{X}\ :\ \limsup_{N\to\infty} \|\Op{T}_{N} f\|_{\mathcal{Y}} = \infty \Big\}\quad\text{and}\\
	\mathcal{D}_{\mathrm{strong}} &= \Big\{ f \in \mathcal{X}\ :\ \lim_{N\to\infty} \|\Op{T}_{N} f\|_{\mathcal{Y}} = \infty \Big\}
\end{split}
\end{equation}
of functions for which weak and strong divergence emerges, respectively.

% ----- LEMMA Divergence Sets -----
\begin{theorem}
\label{thm:DivSets}
Let $\mathcal{X}$ and $\mathcal{Y}$ be two Banach spaces such that $\mathcal{X}$ is continuously embedded in $\mathcal{Y}$, and
let $\Op{T}_{N} : \mathcal{X} \to \mathcal{Y}$ be a sequence of bounded linear operators such that \eqref{equ:ConvOnDenseSubSet} holds for a dense subset $\mathcal{X}_{0} \subset \mathcal{X}$.
For any $M,N \in \NN$ define the set
\begin{equation*}
	D(M,N) := \Big\{ f \in \mathcal{X}\ :\ \| \Op{T}_{N}f \|_{\mathcal{Y}}  > M \Big\}\;.
\end{equation*}
\begin{enumerate}
\item
If $\mathcal{D}_{\mathrm{weak}}$ is non-empty, then for all $M,N_{0} \in \NN$ the set
\begin{equation}
\label{equ:UnionD}
	\bigcup^{\infty}_{N=N_{0}} D(M,N)
\end{equation}
is open and dense in $\mathcal{X}$.
\item
For the divergence sets defined in \eqref{equ:DivSets}, hold
\begin{align}
\label{equ:LemDweak}
	\mathcal{D}_{\mathrm{weak}}
	&= \bigcap^{\infty}_{M=1}\ \limsup_{N\to\infty}\ D(M,N)
	= \bigcap^{\infty}_{M=1}\ \bigcap^{\infty}_{N_{0} = 1}\ \bigcup^{\infty}_{N = N_{0}}\ D(M,N)\\
\label{equ:LemDstrong}	
	\mathcal{D}_{\mathrm{strong}}
	&= \bigcap^{\infty}_{M=1}\ \liminf_{N\to\infty}\ D(M,N)
	= \bigcap^{\infty}_{M=1}\ \bigcup^{\infty}_{N_{0} = 1}\ \bigcap^{\infty}_{N = N_{0}}\ D(M,N)\;.	
\end{align}
\end{enumerate}
\end{theorem}
% ---------------------------------

\begin{remark}
For completeness and for later reference, the straight forward proofs of these statements are provided in the Appendix.
\end{remark}

\begin{remark}
It is easy to see that the operators $\Op{S}_{N} : \PW^{1}_{\pi} \to \B^{\infty}_{\pi}$, associated with the Shannon sampling series and defined in \eqref{equ:ShanSampSer}, satisfy the requirements of Theorem~\ref{thm:DivSets}.
Indeed, Theorem~\ref{thm:standartPW} shows that $\PW^{2}_{\pi}$ is a dense subset of $\PW^{1}_{\pi}$ such that $\lim_{N\to\infty} \|\Op{S}_{N} f_{0} - f_{0}\|_{\infty} = 0$ for all $f_{0} \in \PW^{2}_{\pi}$,
and Theorem~\ref{thm:GlobalDivShannon} implies $\mathcal{D}_{\mathrm{weak}} \neq \varnothing$.
\end{remark}

At a first sight, the structure of both divergence sets seems to be fairly similar.
The only difference is that the inner intersection and union in \eqref{equ:LemDweak} and \eqref{equ:LemDstrong} are interchanged.
However, the different order of these two operations has a distinct consequence.
The first statement of Theorem~\ref{thm:DivSets} shows that the sets \eqref{equ:UnionD} are open and dense subsets of $\mathcal{X}$, provided that $\mathcal{D}_{\mathrm{weak}}$ is non-empty.
Then Theorem~\ref{thm:DivSets} states that $\mathcal{D}_{\mathrm{weak}}$ is a countable intersection of these sets, and
Baire's category theorem implies that a countable intersection of open and dense subsets of a Banach space $\mathcal{X}$ is a set of second category, i.e. a nonmeager set. 
Consequently, if one is able to show that there exists one function in $\mathcal{D}_{\mathrm{weak}}$, representation \eqref{equ:LemDweak} together with the category theorem of Baire implies immediately that $\mathcal{D}_{\mathrm{weak}}$ is nonmeager.

For $\mathcal{D}_{\mathrm{strong}}$ the situation is completely different.
There we have on the right hand side the countable intersection of the open sets $D(M,N)$. 
However, the intersection of open sets is generally no longer open, and it may even be empty.
So the representation \eqref{equ:LemDstrong} gives only little information about the size of $\mathcal{D}_{\mathrm{strong}}$.
If there exists one function $f \in \mathcal{D}_{\mathrm{strong}}$, then it is easy to show (using the same ideas as in part one of the proof of Theorem~\ref{thm:DivSets}) that $\mathcal{D}_{\mathrm{strong}}$ is dense in $\mathcal{X}$.
Nevertheless, even if it is dense, it might be a set of first category, i.e. a meager set.

Finally, we shortly discuss the oscillatory behavior of the Shannon series \cite{BoFa_JApproxT14}. This will give further insight into its divergence behavior. 

% ----- Theorem: Oszillatory -----
\begin{theorem}
\label{thm:oszill}
Let $f \in \PW^{1}_{\pi}$ be a function for which the Shannon sampling series \eqref{equ:ShanSampSer} diverges strongly. Then
\begin{equation}
\label{equ:limMax}
     \lim_{N\to\infty} \left( \max_{t\in\RN}\ (\Op{S}_{N}f)(t) \right) = + \infty 
\end{equation}	
and
\begin{equation*}	
	\lim_{N\to\infty} \left( \min_{t\in\RN}\ (\Op{S}_{N}f)(t) \right) = - \infty \;.
\end{equation*}
\end{theorem}

This result not only implies the statement of Theorem~\ref{thm:StrongDiv} but it additionally shows the oscillatory behavior of the Shannon series and its unlimited growth as $N$ tends to infinity.
To the best of our knowledge, Theorem~\ref{thm:oszill} is the only example which proves the strong oscillatory behavior of a practically relevant reconstruction method, and we will also see in Sec.~\ref{sec:ConvOverSamp} that non-uniform sampling series diverge strongly (cf. Theorem~\ref{thm:DivNUS} below and the corresponding discussion).

We close this section with two examples which illustrate that there are many more problems where the question of strong divergence is of importance.

% ----- Example - Erdös -----
\begin{example}[Lagrange interpolation on Chebyshev nodes]
In 1941, \emph{Paul Erd{\H o}s} tried to show a behavior like \eqref{equ:limMax} for the Lagrange interpolation on Chebyshev nodes.
In \cite{Erdoes41}, he claimed that a statement like \eqref{equ:limMax} holds for the Lagrange interpolation of continuous functions.
However, in \cite{Erdoes43} he observed that his proof was erroneous, and he was not able to present a correct proof.
He presented a result equivalent to Theorem~\ref{thm:StrongDiv}, and
it seems that the original problem is still open.
\end{example}
% ---------------------------

% ----- Example - Hilbert Trafo -----
\begin{example}[Calculation of the Hilbert transform]
For any function $f \in L^1([-\pi,\pi])$ the \emph{Hilbert transform}\index{Hilbert transform} $\Op{H}$ is defined by
\begin{equation}
\label{equ:HilbertTransform}
	(\Op{H}f)(t)
	= \lim_{\epsilon \to 0} \frac{1}{2\pi} \int_{\epsilon \leq |\tau| \leq \pi} \frac{f(t+\tau)}{\tan(\tau/2)}\, \d\tau\;.
\end{equation}
This operation plays a very important roll in different areas of communications, control theory, physics and signal processing \cite{Gabor_ComTheory_1946,PB_Book_AdvTopics,Vakman_HTrafo72}.
In practical applications, $\Op{H}f$ has to be determined based on discrete samples $\{f(t_{n})\}^{N}_{n=-N}$ of $f$.
To this end, let $\{\Op{T}_{N}\}_{N\in\NN}$ any sequence of linear operators which determines an approximation of $\Op{H}f$ from the samples of $\{f(t_{n})\}^{N}_{n=-N}$ of $f$.
It was shown in \cite{Boche_Pohl_IEEE_IT_InterpolatedData} that for any such operator sequence, there exists a function $f_{0} \in \mathcal{B} := \{f \in \mathcal{C}([-\pi,\pi]) : \Op{H}f \in \mathcal{C}([-\pi,\pi])\}$ such that
\begin{equation*}
	\limsup_{N\to\infty} \|\Op{T}_{N} f_{0}\|_{\infty} = \infty\;.
\end{equation*}
In other words, any (non-adaptive) linear method which determines the Hilbert transform from a discrete set of sampled values diverges weakly.
Here, it also would be interesting to investigate whether the questions Q-1 has positive answers or not, i.e. whether there exist adaptive methods to approximate the Hilbert transform from interpolated data.
\end{example}
% -----------------------------------

% ==============================================
\subsection{Convergence for Oversampling}
\label{sec:ConvOverSamp}

So far we saw that the Shannon sampling series diverges strongly on $\PW^{1}_{\pi}$.
However, applying non-uniform sampling patterns and increasing the sampling rate induces additional degrees of freedom, which might give a better convergence behavior.
The question is whether non-uniform sampling resolves the divergence problems observed for uniform sampling.

We start our investigations by showing that the result of Brown (Theorem~\ref{thm:Brown}) on the local uniform approximation behavior of the Shannon sampling series can be extended to non-uniform sampling series, provided that the sampling pattern is equal to the zero set of a sine-type function \cite{BoMo_SigPro10}.

\begin{theorem}
\label{thm:lokConvNUS}
Let $\Lambda = \{\lambda_{n}\}_{n\in\ZN}$ be the zeros set of a sine-type function $\varphi$, let $\{\varphi_{n}\}_{n\in\ZN}$ be the corresponding interpolation kernels, defined in \eqref{equ:Kernels}, and let $\Op{A}_{N} : \PW^{1}_{\pi} \to \B^{\infty}_{\pi}$ be defined as in \eqref{equ:NonUnifSS}.
Then for every $T > 0$, one has
\begin{equation*}
	\lim_{N\to\infty} \left( \max_{t\in [-T,T]} \left| f(t) - (\Op{A}_{N}f)(t) \right| \right) = 0
	\qquad\text{for all}\ f\in\PW^{1}_{\pi}\;.
\end{equation*} 
\end{theorem}

So if sampling patterns derived from sine-type functions are used, then we do not need oversampling to obtain local convergence for all $f \in \PW^{1}_{\pi}$.
It seems natural to ask whether we can apply other sampling patterns to achieve local convergence of $\Op{A}_N f$.
However, we believe that Theorem~\ref{thm:lokConvNUS} is sharp with respect to the chosen sampling pattern. If more general sampling patterns are used, the sampling series $\Op{A}_{N}f$ may no longer converge to $f$. More precisely, we believe that the following statement is true.

\begin{conjecture}
There exist a complete interpolating sequence $\Lambda = \{\lambda_{n}\}_{n\in\ZN}$ with generating function $\varphi$, corresponding interpolation kernels $\{\varphi_{n}\}_{n\in\ZN}$, a point $t_{*} \in \RN$, and a function $f_{*} \in \PW^{1}_{\pi}$ such that
\begin{equation*}
	\limsup_{N\to\infty} \big|(\Op{A}_{N} f_{*})(t_{*})\big|
	= \limsup_{N\to\infty} \left|\sum^{N}_{n=-N} f_{*}(\lambda_{n})\, \varphi_{n}(t_{*})\right|
	= + \infty\;.
\end{equation*}
\end{conjecture}

Next, we ask whether the non-uniform sampling series $\Op{A}_{N} f$ even converges globally uniformly on $\RN$.
It turns out that the answer is negative.
More precisely, one can even show that the non-uniform sampling series $\Op{A}_{N} : \PW^{1}_{\pi} \to \B^{\infty}_{\pi}$, given in \eqref{equ:NonUnifSS}, \emph{diverges strongly}.
To prove this statement, \cite{BoFa_JApproxT14} used non-uniform sampling patterns derived from a special type of sine-type function:
For any $g \in \PW^{1}_{\pi}$, we define the function
\begin{equation}
\label{equ:SineWaveCross}
	\varphi(z) = A\, \sin(\pi z) - g(z)\;,
	\qquad z\in\CN
\end{equation}
with a constant $A > \|g\|_{\PW^{1}_{\pi}} \geq \|g\|_{\infty}$. 
This is a sine-type function and we say that $\varphi$ is determined by the \emph{sine wave crossings}\index{Sine wave crossings} of $g$.
Such functions are used in sampling theory and communications \cite{BarDavid_74,Piwnicki_83} by methods which try to reconstruct the signal from its sine wave crossings.

\begin{theorem}
\label{thm:DivNUS}
Let $\varphi$ be a sine-type function of the form \eqref{equ:SineWaveCross} with zero set $\Lambda = \{\lambda_{n}\}_{n\in\ZN}$ and let $\{ \varphi_{n} \}_{n\in\ZN}$ be the corresponding interpolation kernels \eqref{equ:Kernels}.
Then the non-uniform sampling series \eqref{equ:NonUnifSS} diverges strongly, i.e.
there exists a function $f \in \PW^{1}_{\pi}$ such that
\begin{equation*}
	\lim_{N\to\infty} \|\Op{A}_{N}f\|_{\infty} 
	= \lim_{N\to\infty} \left( \max_{t\in\RN} \left| \sum^{N}_{n=-N} f(\lambda_{n})\, \varphi_{n}(t) \right| \right) = \infty\;.
\end{equation*}
\end{theorem}

So a non-uniform sampling series alone gives no improvement with respect to the global uniform convergence as compared to the uniform sampling considered in Theorem~\ref{thm:StrongDiv}.
Both the uniform and the non-uniform sampling series diverge strongly on $\PW^{1}_{\pi}$.
Note that the previous theorem was formulated only for sampling patterns arising from the zero set of a sine-type function of the form \eqref{equ:SineWaveCross}.
This is only a small subset of all complete interpolating sequences.
Nevertheless, there is strong evidence that Theorem~\ref{thm:DivNUS} also holds for arbitrary complete interpolating sequences.
This gives the following conjecture \cite{BoFa_JApproxT14}.

\begin{conjecture}
\label{con:Conj1}
Let $\Lambda = \{\lambda_{n}\}_{n\in\ZN}$ be an arbitrary complete interpolating sequence with generator $\varphi$ and corresponding interpolation kernels \eqref{equ:Kernels}.
Then there exists an $f \in \PW^{1}_{\pi}$ such that
\begin{equation*}
	\lim_{N\to\infty} \|\Op{A}_{N}f\|_{\infty} 
	= \lim_{N\to\infty} \left( \max_{t\in\RN} \left| \sum^{N}_{n=-N} f(\lambda_{n})\, \varphi_{n}(t) \right| \right) = \infty\;.
\end{equation*}
\end{conjecture}

Non-uniform sampling pattern alone does not resolve the divergence problem of the sampling series on $\PW^{1}_{\pi}$. 
Since $\PW^{1}_{\pi} \subset \B^{\infty}_{\pi}$, the same statement holds for $\B^{\infty}_{\pi}$.
Next we want to investigate whether oversampling improves the global convergence behavior of the non-uniform sampling series \eqref{equ:NonUnifSS}.
This is done for the largest signal spaces $\B^{\infty}_{\pi}$.

Our first theorem in this direction, taken from \cite{MoBo_SigPro10}, shows that if oversampling is applied, then the result of Brown (Theorem~\ref{thm:Brown}) on the local uniform convergence on $\PW^{1}_{\pi}$ can be extended to the larger space $\B^{\infty}_{\pi}$ and to non-uniform sampling series of the form \eqref{equ:NonUnifSS}:

\begin{theorem}
\label{thm:Binfty}
Let $\Lambda = \{\lambda_{n}\}_{n\in\ZN}$ be the zero set of a sine-type function $\varphi$, and let $\{ \varphi_{n} \}_{n\in\ZN}$ be defined as in \eqref{equ:Kernels}.
Then for every $T > 0$ and any $0 < \beta < 1$, we have
\begin{equation*}
	\lim_{N\to\infty} \max_{t \in [-T,T]} \left| f(t) - (\Op{A}_{N} f)(t) \right| = 0
	\quad\text{for all}\ f \in \B^{\infty}_{\beta\pi}
\end{equation*}
where $\Op{A}_{N}$ is defined in \eqref{equ:NonUnifSS}.
\end{theorem}

\begin{remark}
Note that this theorem allows sampling patterns from arbitrary sine-type functions.
The oversampling is expressed by the fact that the above result holds only for functions in $\B^{\infty}_{\beta\pi}$ with $\beta < 1$, i.e. for functions with bandwidth $\beta\pi < \pi$.
\end{remark}

\begin{remark}
If no oversampling is applied, i.e. if $\beta = 1$, then the above result is not true, in general.
Then one can only show \cite{MoBo_SigPro10} that the approximation error remains locally bounded, i.e. $\sup_{N\in\NN} \max_{t \in [-T,T]} |f(t) - (\Op{A}_{N} f)(t)| \leq C\, \|f\|_{\infty}$ for all $f \in \B^{\infty}_{\pi}$.
\end{remark}

The question is whether we can also have uniform convergence on the entire real axis. So what happens if we let $T$ go to infinity in Theorem~\ref{thm:Binfty}?
The answer is given by the next theorem \cite{MoBo_SigPro10}. It shows that we only have uniform convergence on $\RN$ for the subset $\B^{\infty}_{\beta\pi,0}$ of all $f \in \B^{\infty}_{\beta \pi}$ which vanish at infinity.
However, in general, the approximation error remains uniformly bounded for every $f \in \B^{\infty}_{\beta\pi}$.

\begin{theorem}
\label{thm:ConvergBinfty}
Let $\Lambda = \{\lambda_{n}\}_{n\in\ZN}$ be the zero set of a sine-type function $\varphi$, and let $\Op{A}_{N}$ be defined as in \eqref{equ:NonUnifSS} with interpolation kernels $\{ \varphi_{n} \}_{n\in\ZN}$ given in \eqref{equ:Kernels}.
Then for any $0 < \beta < 1$, we have
\begin{equation*}
	\lim_{N\to\infty} \max_{t\in\RN} \left| f(t) - (\Op{A}_{N} f)(t)\right| = 0
	\qquad\text{for all}\ f\in \B^{\infty}_{\beta\pi,0}\;,
\end{equation*}
and there exists a constant $C>0$ such that
\begin{equation*}
	\lim_{N\to\infty} \max_{t\in\RN} \left| f(t) - (\Op{A}_{N} f)(t)\right| \leq C\, \|f\|_{\infty}
	\qquad\text{for all}\ f\in \B^{\infty}_{\beta\pi}\;.
\end{equation*}
\end{theorem}

Since $\PW^{1}_{\sigma} \subset \B^{\infty}_{\sigma,0}$, Theorem~\ref{thm:ConvergBinfty} includes in particular the following corollary \cite{BoMo_SigPro10} on the global uniform convergence of $\Op{A}_{N}$ on $\PW^{1}_{\beta\pi}$.

\begin{corollary}
\label{cor:ConvPW1oversamp}
Let $\varphi$ be a sine-type function with zero set $\Lambda = \{\lambda_{n}\}_{n\in\ZN}$ and let $\{ \varphi_{n} \}_{n\in\ZN}$ be the corresponding interpolation kernels \eqref{equ:Kernels}.
Then for every $f \in \PW^{1}_{\beta\pi}$ with $0<\beta<1$ holds
\begin{equation*}
	\lim_{N\to\infty} \|f - \Op{A}_{N} f\|_{\infty}
	= \lim_{N\to\infty} \left( \max_{t\in\RN} \left| f(t) - \sum^{N}_{n=-N} f(\lambda_{n})\, \varphi_{n}(t) \right| \right)
	= 0\;.
\end{equation*}
\end{corollary}

So with oversampling, i.e. for functions in $\PW^{1}_{\beta\pi}$ with $\beta < 1$, the sampling series \eqref{equ:NonUnifSS} converges uniformly on the entire real axis.
This result can not be extended to the larger signal space $\B^{\infty}_{\beta\pi}$. For these functions, the approximation error is only bounded in general, but does not go to zero as $N$ tends to infinity.
This even holds for any arbitrary large oversampling factor $1/\beta$.

% ============================================================================================
% ========== Sampling-Based Signal Processing ================================================
% ============================================================================================
\section{Sampling-Based Signal Processing}
\label{sec:SampBasedSP}

Up to now, our discussion was based on the goal to reconstruct a certain function, say $f\in\PW^{1}_{\pi}$, from its values $\{f(\lambda_n)\}_{n\in\ZN}$ at the sampling points $\{\lambda_{n}\}_{n\in\ZN}$.
However, in applications one is often not interested in $f$ itself, but in some processed version of $f$, i.e. one wants to determine $g = \Op{H} f$ where $\Op{H} : \PW^{1}_{\pi} \to \PW^{1}_{\pi}$ is a certain linear system, for example the Hilbert transform \eqref{equ:HilbertTransform} or the derivation operator $f(t) \mapsto \d f/\d t$ \cite{Butzer_Entropy12}.
Since signals in the physical world are usually analog, the system $\Op{H}$ is often described in the analog domain. Then for a given function $f$, it would be easy determining $g = \Op{H} f$. 
However, if only samples $\{f(\lambda_{n})\}_{n\in\ZN}$ of $f$ are available, then it seems to be more desirable to implement the system $\Op{H}$ directly in the digital domain, based on the signal samples $\{f(\lambda_n)\}_{n\in\ZN}$.
Thus, we look for a mapping $\Op{H_{D}} : \{f(\lambda_{n})\} \mapsto \Op{H} f$ which determines $g = \Op{H} f$ directly from the available samples $ \{f(\lambda_{n})\}$.
We call $\Op{H_{D}}$ the \emph{digital implementation} of the analog system $\Op{H}$. 

\begin{example}[Sensor Networks]
In a sensor network, many sensors which are distributed non-uniformly in space (and time), measure (i.e. sample) a certain physical quantity (e.g. temperature, pressure, the electric or magnetic field strength, velocity, etc.). For concreteness, assume that we measure temperature.
Then the aim is not necessarily to reconstruct the entire temperature distribution in the observed area, but only to determine, say, the maximum temperature, or the maximum temperature difference.
\end{example}

The interesting question now is whether it is possible to find for a given analog system $\Op{H}$, a digital implementation $\Op{H_{D}}$.
The answer depends strongly on the system $\Op{H}$ under consideration. Here we only investigate this problem for a fairly simple but very important class of mappings $\Op{H}$, namely
for stable linear, time-invariant systems $\Op{H} : \PW^{1}_{\pi} \to \PW^{1}_{\pi}$.

% ========== LTI Systems ==========
\subsection{Linear Time-Invariant Systems}

In our context, a \emph{linear system} is always a linear operator $\Op{H} : \PW^{1}_{\pi} \to \PW^{1}_{\pi}$.
Such a system is called \emph{stable}\index{System!stable}, if $\Op{H}$ is bounded, i.e. if
\begin{equation*}
	\|\Op{H}\|
	= \sup\left\{ \|\Op{H} f\|_{\PW^{1}_{\pi}}\ :\ f\in\PW^{1}_{\pi}\,,\ \|f\|_{\PW^{1}_{\pi}} \leq 1 \right\} < \infty\;,
\end{equation*}
and $\Op{H}$ is said to be \emph{time-invariant} if it commutes with the \emph{translation operator}\index{Operator!translation} $\Op{T}_a : f(t) \mapsto f(t-a)$, i.e. if
$\Op{H} \Op{T}_{a} f = \Op{T}_{a}\Op{H} f$ for all $a\in\RN$ and for every $f\in \PW^{1}_{\pi}$.

It is known that for every stable, linear, time-invariant (LTI) system\index{System!linear, time-invariant (LTI)} $\Op{H} : \PW^{1}_{\pi} \to \PW^{1}_{\pi}$ there exists a unique function $\widehat{h} \in L^{\infty}([-\pi,\pi])$ such that for all $f \in \PW^{1}_{\pi}$
\begin{equation}
\label{equ:LTIsystem}
	(\Op{H} f)(t)
	= \frac{1}{2\pi} \int^{\pi}_{-\pi} \widehat{f}(\omega)\, \widehat{h}(\omega)\, \E^{\I\omega t}\, \d\omega\;,
	\quad t\in\RN\;
\end{equation}
and with $\|H\| = \|\widehat{h}\|_{\infty}$.
Conversely, every function $\widehat{h} \in L^{\infty}([-\pi,\pi])$, defines by \eqref{equ:LTIsystem}, a stable LTI system $\Op{H}$.
In engineering, the function $\widehat{h}$ is often called the \emph{transfer function}\index{Transfer function} of the LTI system $\Op{H}$, whereas its inverse Fourier transform $h =  \mathcal{F}^{-1}\widehat{h}$ is said to be the \emph{impulse response}\index{Impulse response} of $\Op{H}$.
Since $L^{\infty}([-\pi,\pi]) \subset L^{2}([-\pi,\pi])$, we have that $h \in \PW^{2}_{\pi}$.

% ----- Implementation for PW2
\runinhead{Digital implementation for $\PW^{2}_{\pi}$.}
Now we want to find a digital implementation $\Op{H_{D}}$ for a stable LTI system of $\Op{H}$.
To this end we first consider the situation on the Hilbert space $\PW^{2}_{\pi}$. 
There the obvious way to define $\Op{H_{D}}$ is by applying $\Op{H}$ to $\Op{A}_{N}f$. This yields
\begin{equation}
\label{equ:defHD}
	(\Op{H} \Op{A}_{N} f)(t)
	= \sum^{N}_{n=-N} f(\lambda_{n})\, (\Op{H}\varphi_{n})(t)
	= \sum^{N}_{n=-N} f(\lambda_{n})\, \psi_{n}(t)
	=: (\Op{H}_N f)(t)
\end{equation}
with kernels $\psi_{n} := \Op{H}\varphi_{n} \in \PW^{2}_{\pi}$, for all $n\in\ZN$, and where the sampling set $\Lambda$ is chosen to be a complete interpolating for $\PW^{2}_{\pi}$.
If $\Op{H}$ is a stable system $\PW^{2}_{\pi} \to \PW^{2}_{\pi}$, then it follows from Theorem~\ref{thm:NonUniformSamp} that 
\begin{equation*}
	\|\Op{H} f - \Op{H}_{N} f\|_{\PW^{2}_{\pi}}
	= \|\Op{H} f - \Op{H} \Op{A}_{N} f \|_{\PW^{2}_{\pi}}
	\leq \|\Op{H}\|\, \|f - \Op{A}_{N} f\|_{\PW^{2}_{\pi}} \to 0
\end{equation*}
as $N\to\infty$ for every $f \in \PW^{2}_{\pi}$. Since $\PW^{2}_{\pi}$ is a reproducing kernel Hilbert space, the norm convergence again implies the uniform convergence on $\RN$.

% ----- Implementation for PW1 -----
\runinhead{Digital implementation in $\PW^{1}_{\pi}$.}
Since $\PW^{2}_{\pi}$ is a dense subset of $\PW^{1}_{\pi}$, we may hope that the implementation for $\PW^2_{\pi}$ extends in some sense to $\PW^{1}_{\pi}$.

Let us first consider a very special stable LTI system, namely the identity operator $\Op{H} = \Op{I}_{\PW^{1}_{\pi}}$ on  $\PW^{1}_{\pi}$.
For this particular system, its digital implementation is easily derived. 
Following the above ideas for $\PW^{2}_{\pi}$, its digital implementation is simply $\Op{H}_{N} = \Op{H} \Op{A}_{N} = \Op{A}_{N}$, and
Corollary~\ref{cor:ConvPW1oversamp} implies
\begin{equation*}
	\lim_{N\to\infty} \|\Op{H} f - \Op{H}_{N} f \|_{\infty}
	= \lim_{N\to\infty} \|\Op{H}f - \Op{A}_{N} f \|_{\infty}
	= \lim_{N\to\infty} \max_{t\in\RN} \left| \Op{H}f - \sum^{N}_{n=-N} f(\lambda_{n})\, \varphi_{n}(t) \right|
	= 0
\end{equation*}
for all $f \in \PW^{1}_{\beta\pi}$ with $\beta < 1$, and provided that the sampling set $\Lambda = \{\lambda_{n}\}_{n\in\ZN}$ was chosen to be the zero set of a sine-type function.
So for the identity operator, we found a digital implementation. 
Since Corollary~\ref{cor:ConvPW1oversamp} was used in the above arguments, it is clear that this digital implementation is based on the oversampled input signal $f$.
Theorem~\ref{thm:DivNUS} shows then that if $\Lambda$ is the zero set of sine-type functions of the form \eqref{equ:SineWaveCross}, oversampling is indeed necessary even for the global approximation of the simple LTI  $\Op{H} = \Op{I}_{\PW^{1}_{\pi}}$.
Moreover, if Conjecture~\ref{con:Conj1} turns out to be true, then it would imply that oversampling is necessary for all complete interpolating sequences $\Lambda$.

Our next question is whether the digital implementation \eqref{equ:defHD} converges for any arbitrary stable LTI system $\Op{H}$ on $\PW^{1}_{\beta\pi}$, i.e. whether $\Op{H}_{N} f \to \Op{H} f$ as $N \to \infty$ for all $f \in \PW^{1}_{\beta\pi}$.
In particular, we investigate whether $\Op{H}_{N} f$ converges locally uniformly to $\Op{H} f$ or even globally uniformly as the identity operator.

% ------------------------------------------
% ----- Divergenc of Point Evaluations -----
% ------------------------------------------
\subsection{Sampling via Point Evaluations}

So the digital implementation $\Op{H}_{N}$  of any stable LTI system $\Op{H} : \PW^{1}_{\pi} \to \PW^{1}_{\pi}$ is defined as in \eqref{equ:defHD},
based on a complete interpolating sequence $\Lambda = \{\lambda_{n}\}_{n\in\ZN}$ for $\PW^{2}_{\pi}$ and based on the 
interpolation kernels $\{\varphi_{n}\}_{n\in\ZN}$ given in \eqref{equ:Kernels} with $\varphi$ as in \eqref{equ:DefVarPhi}.

The next theorem taken from \cite{BoMo_Butzer14} shows that there exist stable LTI systems for which such a digital implementation is not possible, even if we allow arbitrarily large oversampling.
More precisely, it shows that there exist stable LTI systems such that the approximation of its digital implementation $\Op{H}_{N}f$ diverges even pointwise for some $f \in \PW^{1}_{\pi}$.

\begin{theorem}
\label{thm:DivSystemImpl}
Let $\Lambda = \{\lambda_{n}\}_{n\in\ZN}$ be a complete interpolating sequence for $\PW^{2}_{\pi}$ and let $\{ \varphi_{n} \}_{n\in\ZN}$ be the interpolation kernels defined in \eqref{equ:Kernels}. 
Let $t\in\RN$ be arbitrary, then there exists a stable LTI system $\Op{H} : \PW^{1}_{\pi} \to \PW^{1}_{\pi}$ such that for every $0 < \beta < 1$ there exists a signal $f \in \PW^{1}_{\beta\pi}$ such that
\begin{equation*}
	\limsup_{N\to\infty} \left|(\Op{H}_{N} f)(t)\right|
	= \limsup_{N\to\infty} \left| \sum^{N}_{n=-n} f(\lambda_{n})\, (\Op{H}\varphi_{n})(t)\right|
	= \infty\;.
\end{equation*}
\end{theorem}

So for any fixed $t\in\RN$ there are stable LTI systems $\Op{H} : \PW^{1}_{\pi} \to \PW^{1}_{\pi}$ such that for every $\beta\in(0,1]$ the corresponding digital approximation $\Op{H}_{N}$ diverges at $t$ for some signals $f \in \PW^{1}_{\beta\pi}$.
But on the other side, since $\Op{H}$ is stable, we have for any $t \in \RN$ and any $f \in \PW^{1}_{\pi}$
\begin{equation*}
	\left| (\Op{H}f)(t) \right|
	\leq \|\Op{H} f\|_{\infty}
	\leq \|\Op{H} f\|_{\PW^{1}_{\pi}}
	\leq \| \Op{H} \|\, \|f\|_{\PW^{1}_{\pi}} < \infty\;.
\end{equation*}
So the divergence observed in Theorem~\ref{thm:DivSystemImpl} is indeed a property of the digital implementation of $\Op{H}$ based on (time domain) samples of $f$ and not a property of the system $\Op{H}$ itself.
Moreover, if it were possible to sample $f$ in the frequency domain, then we could approximate the integral in its analog implementation \eqref{equ:LTIsystem} by its Riemann sum. This sum would converge to $(\Op{H}f)(t)$ for every $t\in\RN$.

Overall, we see that there exists no general answer to the question whether every LTI system $\Op{H} : \PW^{1}_{\pi} \to \PW^{1}_{\pi}$ can be implemented digitally.
Of course there are stable LTI systems which allow such digital implementation.
The identity operator discussed above, is one example of such a system.
However, Theorem~\ref{thm:DivSystemImpl} shows that there exist stable systems for which such a digital implementation is not possible.

We come back to the discussion at the end of Section~\ref{sec:Divergence}, and ask whether question Q-1 or Q-2 may have a positive answer for the approximation operators $\Op{H}_{N}$,
i.e. whether there exist subsequences $\{N_{k}\}_{k\in\NN}$ (dependent on the function $f$, or not) such that $\{\Op{H}_{N_{k}}f\}_{k\in\NN}$ converges to $\Op{H}f$.
To this end, let $\Op{H}$ be a stable LTI system and let $t\in\RN$ be a fixed point. Then $(\Op{H}_{N}f)(t)$ defines a sequence of linear functionals on $\PW^{1}_{\beta\pi}$, for every $\beta\in (0,1]$, with the norm
\begin{equation*}
	\|\Op{H}_{N}\|_{t,\beta} 
	= \sup\left\{ \left| (\Op{H}_{N} f)(t) \right|  :\ f\in \PW^{1}_{\beta\pi}\;,\ \|f\|_{\PW^{1}_{\beta\pi}} \leq 1\right\}\;.
\end{equation*}
Then Theorem~\ref{thm:DivSystemImpl} implies that for every $t\in\RN$ there exists a stable LTI system $\Op{H}$ such that for all $\beta \in (0,1]$
\begin{equation*}
	\limsup_{N\to\infty} \left\| \Op{H}_{N} \right\|_{t,\beta} = +\infty\;.
\end{equation*}
However, since we have no statement for $\liminf_{N\to\infty} \|\Op{H}_{N}\|_{t,\beta}$, we do not know at the moment whether $\|\Op{H}_{N}\|_{t,\beta}$ satisfies an inequality similar to \eqref{equ:LebesgueKonstantSN} for some $t\in\RN$.
If a lower bound like \eqref{equ:LebesgueKonstantSN} were to exist, then question Q-2 would have a negative answer, i.e. no subsequence $\{N_{k}\}_{k\in\NN}$ would exist such that $H_{N_{k}} f$ converges globally uniformly to $\Op{H} f$ for all $f\in\PW^{1}_{\pi}$.
Indeed, we believe that the following statement is true, which would imply a negative answer to Q-2 (see discussion in Section~\ref{sec:Divergence}).

\begin{conjecture}
Let $\Lambda = \{\lambda_{n}\}_{n\in\ZN}$ be a complete interpolating sequence for $\PW^{2}_{\pi}$, and let $t\in\RN$ be arbitrary.
Then there exists a stable LTI system $\Op{H} : \PW^{1}_{\pi} \to \PW^{1}_{\pi}$ such that for every $\beta \in (0,1]$
\begin{equation*}
	\lim_{N\to\infty} \|\Op{H}_{N}\|_{t,\beta} = +\infty\;.
\end{equation*}	
\end{conjecture}

It would also be interesting to investigate question Q-1, i.e. to ask whether the sequence $\{\Op{H}_{N} : \PW^{1}_{\beta\pi} \to \B^{\infty}_{\pi}\}$ diverges strongly.
We believe that this is indeed the case, i.e. we think that the following conjecture is true. 

\begin{conjecture}
Let $\Lambda = \{\lambda_{n}\}_{n\in\ZN}$ be a complete interpolating sequence for $\PW^{2}_{\pi}$.
There exists a stable LTI system $\Op{H} : \PW^{1}_{\pi} \to \PW^{1}_{\pi}$ such that for every $\beta \in (0,1]$ there exists an $f_{\beta} \in \PW^{1}_{\beta\pi}$ for which
\begin{equation*}
	\lim_{N\to\infty} \left\| \Op{H}_{N} f_{\beta} \right\|_{\infty}
	= \lim_{N\to\infty} \left( \max_{t\in\RN} \left| (\Op{H}_{N}f_{\beta})(t) \right| \right) = + \infty\;.
\end{equation*}
\end{conjecture}

\begin{remark}
Note that the LTI system $\Op{H} : \PW^{1}_{\pi} \to \PW^{1}_{\pi}$ for which the approximation $\Op{H}_{N}$ diverges is universal with respect to $\beta$.
In other words, we believe that it is not possible to find a digital implementation of $\Op{H}$, regardless of the amount of oversampling.
\end{remark}

If this conjecture is true, it would exclude the existence of an adaptive algorithm which chooses the approximation sequence $\{N_{k}(f)\}_{k\in\NN}$ subject to the actual function $f$ to approximate the output $\Op{H}f$ of the system $\Op{H}$ from the signal samples $\{f(\lambda_{n}) \}_{n\in\ZN}$.

% -----------------------------------------------
% ----- Generalized Measurement Functionals -----
% -----------------------------------------------
\subsection{Sampling by Generalized Measurement}
\label{sec:GenMeasFunc}

The fundamental concept of digital signal processing is to represent analog (i.e. continuous) signals as a sequence of numbers.
In the previous discussions it was always assumed that the conversion from the analog to the digital domain is based on point evaluations of the analog signal.
Thus the measurement functionals were assumed to be of the form
\begin{equation}
\label{equ:PointEvalu}
	\gamma_{n} : f \mapsto f(\lambda_{n})\;,
	\qquad n\in \ZN
\end{equation}
with a certain sequence $\{\lambda_{n}\}_{n\in\ZN}$ of sampling points.
However, more general measurement methods are possible, which we want to investigate next.

Although we depart from the point evaluations \eqref{equ:PointEvalu}, we still require that our measurements are based on bounded linear functionals on the specific function space.
Again, we first consider the situation on the Hilbert space $\PW^{2}_{\pi}$. 
By the Riesz representation theorem\index{Theorem!Riesz representation}, we know that any bounded linear functional $\gamma_{n} : \PW^{2}_{\pi} \to \CN$ can be written as an inner product with a certain function $s_{n} \in \PW^{2}_{\pi}$, i.e.
\begin{equation}
\label{equ:GEneralMeasure}
	\gamma_{n}(f)
	= \left\langle f,s_{n} \right\rangle_{\PW^{2}_{\pi}}
	= \int^{\infty}_{-\infty} f(t)\, \overline{s_{n}(t)}\, \d t
	= \frac{1}{2\pi}\int^{\pi}_{-\pi} \widehat{f}(\omega)\, \overline{\widehat{s}_{n}(\omega)}\, \d\omega\;,
\end{equation}
where the last equation follows from Parseval's formula, and Cauchy-Schwarz inequality gives immediately $\|\gamma_{n}\| = \|s_{n}\|_{\PW^{2}_{\pi}}$.
In this respect, any generalized sampling process on $\PW^{2}_{\pi}$ is based on a sequence $\{s_{n}\}_{n\in\NN}$ of sampling functions in $\PW^{2}_{\pi}$, which defines by \eqref{equ:GEneralMeasure}, a sequence $\{\gamma_{n}\}_{n\in\NN}$ of measurement functionals.
A stable reconstruction of any $f \in \PW^{2}_{\pi}$ from the samples $\{\gamma_{n}(f)\}_{n\in\NN}$ is possible if $\{s_{n}\}_{n\in\NN}$ is at least a \emph{frame}\index{Frame} \cite{Christensen_Frames,Young_NonHarmonic} for $\PW^{2}_{\pi}$.
Let $\{\sigma_{n}\}_{n\in\NN}$ be the dual frame of $\{s_{n}\}_{n\in\NN}$, then $f$ can be reconstructed from its samples $\{\gamma_{n}(f)\}_{n\in\NN}$ by
\begin{equation*}
	f(t)
	= \lim_{N\to\infty} (\Op{A}_{N} f)(t)
	\qquad\text{where}\qquad
	(\Op{A}_{N} f)(t) = \sum^{N}_{n=1} \gamma_{n}(f)\, \sigma_{n}(t)
\end{equation*} 
and where the sum converges in the norm of $\PW^{2}_{\pi}$ and uniformly on $\RN$.
If $\{s_{n}\}_{n\in\ZN}$ is even an orthonormal basis for $\PW^{2}_{\pi}$, then we simply have $\sigma_{n} = s_{n}$ for all $n\in \ZN$.

% --- EXAMPLE - Point Evaluations ---
\begin{example}
The point evaluations \eqref{equ:PointEvalu} can be written as in \eqref{equ:GEneralMeasure} by choosing $s_{n}$ to be equal to the reproducing kernels $r_{\lambda_{n}}$ of $\PW^{2}_{\pi}$.
Moreover, it is known \cite{Young_NonHarmonic} that $\{s_{n}\}_{n\in\ZN}$ is a Riesz basis for $\PW^{2}_{\pi}$ if and only if $\{\lambda_{n}\}_{n\in\ZN}$ is complete interpolating for $\PW^{2}_{\pi}$.
Note that the measurement functionals associated with the point evaluations are uniformly bounded, because $\|\gamma_{n}\| = \|r_{\lambda_{n}}\|_{\PW^{2}_{\pi}} = 1$ for all $n\in\ZN$.
\end{example}
% -----------------------------------

Now we apply again a stable LTI system $\Op{H}$ to the approximation operator $\Op{A}_{N}$. This gives an approximation of the digital implementation $\Op{H}_{\mathrm{D}}$ of $\Op{H}$
\begin{equation}
\label{equ:HNgeneralized}
	(\Op{H}_N f)(t)
	:= (\Op{H}\Op{A}_{N} f)(t)
	= \sum^{N}_{n=1} \gamma_{n}(f)\, (\Op{H}\sigma_{n})(t)\;.
	%= \sum^{N}_{n=1} \gamma_{n}(f)\, \psi_{n}(t)
\end{equation}
If $\Op{H}$ is a stable LTI system $\PW^{2}_{\pi} \to\PW^{2}_{\pi}$, then it is again easy to see that $\Op{H}_{N} f \to \Op{H} f$ as $N\to\infty$ in the norm of $\PW^{2}_{\pi}$ and uniformly on $\RN$ for every $f \in \PW^{2}_{\pi}$.

Now we consider the approximation operator \eqref{equ:HNgeneralized} on $\PW^{1}_{\pi}$.
To this end, $\{\gamma_{n}\}_{n\in\NN}$ has to be a sequence of bounded linear functionals on $\PW^{1}_{\pi}$.
It is known that every bounded linear functional $\gamma_{n} : \PW^{1}_{\pi} \to \CN$ has the form \eqref{equ:GEneralMeasure} but with a function $\widehat{s}_{n} \in L^{\infty}([-\pi,\pi])$ and such that $\|\gamma_{n}\| = \|\widehat{s}_{n}\|_{\infty}$.
As in the case of point evaluations on $\PW^{2}_{\pi}$, we require that all measurement functionals are uniformly bounded, i.e. we require that there exists a positive constant $C_{\gamma}$ such that
\begin{equation}
\label{equ:uniformBounded}
	\|\gamma_{n}\| = \|\widehat{s}_{n}\|_{\infty} \leq C_{\gamma}
	\quad\text{for all}\ n\in \NN\;.
\end{equation}
The question is whether we can find a frame $\{s_{n}\}_{n\in\NN}$ for $\PW^{2}_{\pi}$ such that the series \eqref{equ:HNgeneralized} converges to $\Op{H}f$ for any stable LTI system $\Op{H} : \PW^{1}_{\pi} \to \PW^{1}_{\pi}$ and for every $f\in\PW^{1}_{\pi}$.
The answer is affirmative, provided oversampling is applied.
Moreover, appropriate measurement functionals $\{\gamma_{n}\}_{n\in\NN}$ are generated by an orthonormal sequence $\{s_{n}\}_{n\in\NN}$ in $\PW^{2}_{\pi}$.
More precisely, the following statement can be proved \cite{BoMo_Butzer14}.

% ----- THEOREM - Generalized Measurements -----
\begin{theorem}
\label{thm:GeneralizedMFkt}
Let $0 < \beta < 1$ be arbitrary.
There exists an orthonormal basis $\{s_{n}\}_{n\in\NN}$ for $\PW^{2}_{\pi}$ with the associated measurement functionals \eqref{equ:GEneralMeasure} which satisfy \eqref{equ:uniformBounded}
such that for all stable LTI systems $\Op{H} : \PW^{1}_{\pi} \to \PW^{1}_{\pi}$ and for all $f \in \PW^{1}_{\beta\pi}$
\begin{equation}
\label{equ:ConvDigImpl}
	\lim_{N\to\infty} \|\Op{H} f - \Op{H}_{N} f\|_{\infty}
	= \lim_{N\to\infty} \left(
	\sup_{t\in\RN} \left| (\Op{H}f)(t) - \sum^{N}_{n=1} \gamma_{n}(f) (\Op{H} s_{n})(t) \right|
	\right) = 0\;.
\end{equation}
Moreover, there exists a constant $C_{s}$ such that
\begin{equation*}
	\|\Op{H}_{N} f\|_{\infty}
	= 
	\sup_{t\in\RN}
	\left| \sum^{N}_{n=1} \gamma_{n}(f)\, (\Op{H}s_{n})(t) \right|
	\leq C_{s}\, \|\Op{H}\|\, \|f\|_{\PW^{1}_{\pi}}
	\quad\text{for all}\ f\in\PW^{1}_{\beta\pi}\;.
\end{equation*}
\end{theorem}
% ----------------------------------------------

This theorem shows that there exists a set $\{\gamma_{n}\}_{n\in\ZN}$ of generalized measurement functionals such that, in connection with oversampling, every stable LTI system on $\PW^{1}_{\pi}$ possesses a digital implementation. The measurement functionals $\gamma_{n}$ are defined via \eqref{equ:GEneralMeasure} by a specific orthonormal basis $\{s_{n}\}_{n\in\ZN}$ for $\PW^{2}_{\pi}$.
It should be noted that this orthonormal basis depends on $\beta$, i.e. on the amount of oversampling.
Theorem~\ref{thm:DivSystemImpl} shows that $\{s_{n}\}$ can not be a sequence of reproducing kernels because this would yield point evaluations as measurement functionals.
The proof of Theorem~\ref{thm:GeneralizedMFkt} in \cite{BoMo_Butzer14} is constructive in the sense that it provides an explicit construction of an orthonormal basis $\{s_{n}\}_{n\in\NN}$ such that \eqref{equ:ConvDigImpl} holds. This construction is based on the \emph{Olevskii system}\index{Olevskii system} \cite{Olevskii_66} which is an orthonormal basis for $\mathcal{C}([0,1])$.

In conclusion, this section showed that o $\PW^{1}_{\pi}$, a digital implementation of a stable LTI system can only be guaranteed if generalized measurement functionals with oversampling are used.
If the data acquisition is based on simple point evaluations, then there are stable LTI systems which possess no digital implementation.
This demonstrates in particular the limitation of digital implementations based on measurements from sensor networks, because the sampling process in such a network is basically a point evaluation at the particular sensor position.

% ============================================================================
% ========== Phase Retrievaal ================================================
% ============================================================================
\section{Signal Recovery from Amplitude Samples}
\label{sec:Phase Retrieval}

Sampling theory as discussed in the previous sections is based on signal samples taken by a sequence of linear functionals $\{\gamma_{n}\}_{n\in\ZN}$.
Then the reconstruction method is linear, namely a simple interpolation series of the form \eqref{equ:NonUnifSS} or \eqref{equ:SampSerR}.
If the measurement functionals are non-linear, signal recovery will become more involved and in particular non-linear, in general.

This section discusses a particular case of non-linear measurements which is of considerable interest in many applications.
Assume again that $\{\gamma_{n}\}$ is a set of linear functionals on our signal space and $\{\gamma_{n}(f)\}_{n\in\ZN}$ is the sequence of complex valued samples of a signal $f$.
In many different applications it is not possible to measure the magnitude and the phase of $\gamma_{n}(f)$, but only the squared modulus $|\gamma_{n}(f)|^{2}$.
In this case, the sampling operator $f \mapsto \{|\gamma_{n}(f)|^{2}\}_{n\in\ZN}$ is non-linear.
However, the non-linearity arises only due to the intensity measurement $|\cdot|^{2}$ but it is often possible to design the linear functionals $\{ \gamma_{n} \}_{n\in\ZN}$ by an appropriate measurement setup.
The interesting question is now whether $f$ can be reconstructed from the intensity samples $\{ |\gamma_{n}(f)|^{2} \}_{n\in\ZN}$ and how we have to choose the functionals $\{\gamma_{n}\}_{n\in\ZN}$ such that signal recovery becomes possible.

The described problem, also known as \emph{phase retrieval}\index{Phase retrieval}, arises especially in optics, where, because of the short wavelength, only the intensity of the electromagnetic wave can be measured, but not its actual phase.
Applications where such problems appear range from X-ray crystallography \cite{Millane_90, Miao_XRay_08}, astronomical imaging \cite{Fienup_93}, radar, \cite{Jaming_Radar_JFAA99}, speech processing \cite{Balan_RecWithoutPhase_06} to quantum tomography \cite{Finkelstein_QuantumCom04}, to mention only some.

Phase retrieval for signals from finite dimensional spaces $\CN^{N}$ were considered extensively in the last years. Now there exists necessary and sufficient conditions on the number of samples as well as different recovery algorithms ranging from algebraic methods to algorithms based on convex optimization \cite{Balan_RecWithoutPhase_06,Balan_Painless_09,bodmann_StablePR2014,CandesEldar_PhaseRetrieval,Candes_PhaseLift,PYB_STIP14}.
For infinite dimensional signal spaces, only a few results exist up to now. 
Nevertheless, it seems natural to ask whether it is possible to obtain results for bandlimited signals which are similar to the sampling series considered in the previous section, but which are based on the sampled amplitude only.

Since only the amplitudes of the signal samples are available, some oversampling has to be used to compensate this information loss.
However, several questions arise: How much oversampling is necessary and what are sufficient conditions on the measurement functionals $\{\gamma_{n}\}$ such that signal recovery can be guaranteed?
In particular, can we recover every signal from the amplitudes of point evaluations $\gamma_{n}(f) = f(\lambda_{n})$ or do we need generalized measurement functionals (as discussed in Sec.~\ref{sec:GenMeasFunc})?

Before we start our discussion, we want to mention that in the considered situation, signal recovery will only be possible up to an unknown global phase factor.
To see this, let $\{ \gamma_{n} \}_{n\in\ZN}$ be the set of linear measurement functionals.
Then it is not possible to recover $f$ perfectly from the intensity measurements $\{ | \gamma_{n}(f) |^{2} \}$, but only up to a unitary constant.
Because if $\widetilde{f}(z) = c\, f(z)$, where $c$ is a unitary constant, then both functions $f$ and $\widetilde{f}$ will give the same measurements, i.e. $|\gamma_{n}(\widetilde{f})|^{2} = |\gamma_{n}(f)|^{2}$ for all $n$.
Consequently, we consider here only signal recovery up to a global unitary constant, which is sufficient in most applications.

% --------------------------------------------
\runinhead{Real valued bandlimited functions}

For real valued bandlimited signals there exists a remarkable result \cite{Thakur_2011} in the spirit of classical Shannon sampling theory.
It shows that any signal in the Bernstein spaces $\B^{p}_{\pi}$ with $0 < p \leq \infty$ can be reconstructed from amplitude samples taken at an average rate of at least twice the Nyquist rate.
The result in \cite{Thakur_2011} implies in particular the following statement.

\begin{theorem}
\label{thm:Thakur}
Let $\Lambda = \{\lambda_{n}\}_{n\in\ZN}$ be the zero set of a sine-type function $\varphi$,
and for $1 \leq p \leq \infty$ let $f \in \B^{p}_{\pi/2}$ be real valued on $\RN$.
Then $f$ can uniquely be determined from the samples $c_{n} = |\gamma_{n}(f)|^2 = |f(\lambda_n)|^2$, $n\in\ZN$, up to a sign factor.
\end{theorem}

The proof of Theorem~\ref{thm:Thakur} in \cite{Thakur_2011} also provides a reconstruction algorithm.
It is noteworthy that in the case of real valued functions, point evaluations are sufficient as measurement functionals $\gamma_{n}$.
Unfortunately, the technique used to prove Theorem~\ref{thm:Thakur} in \cite{Thakur_2011} can not easily be extended to the complex valued functions.

% --------------------------------------------
\runinhead{Complex valued bandlimited functions}

In the complex case, simple point evaluations does not seem to be sufficient, but rather very specific measurement functionals have to be chosen.
In \cite{Yang_SampTA13}, measurement functionals $\{\gamma_{n}\}$ for phase retrieval in $\B^{2}_{\pi}$ were proposed, which consist of linear combinations of point evaluations at specific sampling points.
This approach was later extended to all Bernstein spaces $\B^{p}_{\pi}$ with $1 < p < \infty$ in \cite{PYB_JFAA14}.
More precisely, let $f \in \B^{p}_{\pi}$ and let $K\geq 2$ be arbitrary integer, then the measurement functionals proposed in \cite{Yang_SampTA13} are given by 
\begin{equation}
\label{equ:MeasureFkt}
	\gamma_{n,m}(f) = \sum^{K}_{k=1} \overline{\alpha_{k,m}}\, f(n \beta + \lambda_{k})\;,
	\qquad n\in \ZN\;,\ m=1,2,\dots,K^{2}
\end{equation}
where the constant $\beta > 0$, the complex numbers $\{\lambda_{k}\}^{K}_{k=1}$, and the complex coefficients $\{\alpha_{k,m}\}$ are chosen in a very specific way.
To formulate sufficient conditions on the functionals \eqref{equ:MeasureFkt} such that signal recovery is possible, we define the $\CN^{K}$-vectors 
\begin{equation*}
	a_{m} = (\alpha_{1,m} , \dots, \alpha_{K,M})^{\T}\;,
	\qquad m=1,\dots,K^{2}\;.
\end{equation*}
Therewith the requirements on the functionals \eqref{equ:MeasureFkt} can be formulated as follows:

% ----- Recovery Condition -----
\begin{definition}[Recovery condition]
\label{def:RevCon}
We say that the measurement functionals \eqref{equ:MeasureFkt} satisfy the \emph{recovery condition}, if 
\begin{eqnarray*}
	&\text{1)}& \quad \lambda_{K} = \lambda_{1} + \beta\\[1ex]
	&\text{2)}& \quad \Lambda = \{ n\beta + \lambda_{k} : n\in\ZN\;, k=1,\dots, K-1\}\
	\text{is the zero set of a sine-type function}\\
	&\text{3)}& \quad \{a_{m}\}^{K^2}_{m=1}\ \text{forms a $2$-uniform $K^{2}/K$ tight frame for $\CN^{K}$}\;. 
\end{eqnarray*}
\end{definition}
% ------------------------------

\begin{remark}
The particular form of the measurement functionals arises from a concrete measurement setup for a particular phase retrieval problem.
Therefore, there exists a fairly simple practical implementation of these functionals. 
We also remark that the conditions on the functional can be slightly weakened, c.f. \cite{Yang_SampTA13,PYB_JFAA14}.
\end{remark}

\begin{remark}
\label{rem:SamSyst}
It is fairly easy to find coefficients for the measurement functionals \eqref{equ:MeasureFkt} such that the recovery condition is satisfied. 
In particular, constructions for $2$-uniform tight frames can be found in \cite{Zauner_Quantendesigns}.
To get appropriate $\lambda_{k}$ one can choose $\Lambda = \{\lambda_{n} : n\in\ZN\}$ as the zeros set of the sine-type function $\varphi(z) = sin(\pi z)$.
Then $\lambda_{n} = n$, $n = 1,\dots, K$ and $\beta = K - 1$.  
\end{remark}

\begin{theorem}
\label{thm:JFAA}
For any $K\geq2$ let $\{\gamma_{n,m}\}$ be the measurement functionals given in \eqref{equ:MeasureFkt} such that they satisfy the recovery condition.
Let $1 < p < \infty$ and set
\begin{equation*}
	\mathcal{R}^{p}_{\pi} := \{ f \in \B^{p}_{\pi} : f(n\beta + \lambda_{1}) \neq 0\ \text{for all}\ n\in \ZN\}\;.
\end{equation*}
Then every $f \in \mathcal{R}^{p}_{\pi}$ can be recovered from the amplitude measurements
\begin{equation*}
	c_{n,m} = |\gamma_{n,m}(f)|^{2}\;,
	\qquad n\in\ZN\;,\ m=1,\dots,K^{2}
\end{equation*}
up to a global unitary factor.
\end{theorem}

The recovery procedure which belongs to Theorem~\ref{thm:JFAA} consists basically of a two-step procedure.
\begin{enumerate}
\item
In the first step, one determines all values $f(n\beta + \lambda_{k})$ from the amplitude measurements $|\gamma_{n,m}(f)|^{2}$ using ideas and algorithms from finite dimensional phase retrieval.
\item
Since $\Lambda = \{ n\beta + \lambda_{k} : n\in\ZN\;, k=1,\dots, K-1\}$ is the zero set of a sine-type function, we can use the sampling series discussed in Section~\ref{sec:GUConv} to recover $f$ from its samples at the sampling set $\Lambda$.
\end{enumerate}

The first step in this recovery procedure relies only on an appropriate choice of the coefficients $\{\alpha_{k,m}\}$, whereas the second step only relies on an appropriate choice of the sampling set $\Lambda$.
For this reason, it is also easy to extend Theorem~\ref{thm:JFAA} to other signal spaces.
For example, applying Corollary~\ref{cor:ConvPW1oversamp} in the second step of the recovery procedure, we immediately obtain a phase retrieval result for functions in $\PW^{1}_{\pi}$.

\begin{corollary}
\label{cor:JFAA}
For any $K\geq2$ let $\{\gamma_{n,m}\}$ be the measurement functionals given in \eqref{equ:MeasureFkt} such that they satisfy the recovery condition.
Let $0 < \beta < 1$ and set
\begin{equation*}
	\mathcal{P}^{1}_{\pi} := \{ f \in \PW^{1}_{\pi} : f(n\beta + \lambda_{1}) \neq 0\ \text{for all}\ n\in \ZN\}\;.
\end{equation*}
Then every $f \in \mathcal{P}^{1}_{\pi}$ can be recovered from the amplitude measurements
$c_{n,m} = |\gamma_{n,m}(f)|^{2}$, $n\in\ZN$, $m=1,\dots,K^{2}$ up to a global unitary factor.
\end{corollary}

The overall sampling rate in Theorem~\ref{thm:JFAA} is mainly determined by the constant $K$, which can be an arbitrary natural number $K\geq 2$. 
We see from \eqref{equ:MeasureFkt} that we apply $K^{2}$ measurement functional $\gamma_{n,m}$ in every interval of length $\beta$, i.e. the overall sampling rate becomes
$R = K^{2}/\beta$, and $\beta$ has to be chosen such that the sequence $\{n\beta + \lambda_{k} : n\in\ZN, k=1,\dots,K-1\}$ is the zero set of a sine-type function. This implies (see also Remark~\ref{rem:SamSyst}) that $\beta \leq K-1$. Therefore, the overall sampling rate has to be at least $R \geq K^{2}/(K-1) \geq 4$, where $R = 4$ is achieved for $K=2$. So we have found a sufficient condition on the sampling rate.

In Theorem~\ref{thm:JFAA}, functions $f \in \B^{p}_{\pi}$ which have a zero in the set $\{n\beta + \lambda_1 : n\in\ZN \}$ cannot be recovered.
However, on the one hand, it is not hard to see that the set of these functions is fairly small, namely it is a set of first category.
On the other hand, it was also shown in \cite{PYB_JFAA14} that this restriction on the recoverable functions can be avoided if the desired signal $f$ is preprocessed in a specific way, namely by adding a known sine-type function $u$ prior to the amplitude measurements.

\begin{theorem}
\label{thm:JFAA_Binfty}
Let $A_{\mathrm{max}}>0$ be arbitrary, and let $0 < \beta < \beta_{1} < 1$. For any $1 \leq p \leq \infty$ set
\begin{equation*}
	\mathcal{S}^{p} := \{f \in \B^{p}_{\beta\pi} : \|f\|_{\B^{p}_{\beta\pi}} \leq A_{\mathrm{max}} \}\;.
\end{equation*}
Then there exists a sine-type function $u \in \B^{\infty}_{\beta_{1}\pi}$ and a sequence of measurement functionals of the form \eqref{equ:MeasureFkt} which satisfy the recovery condition such that every $f \in \mathcal{S}^{p}$ can be recovered from the amplitude measurements
\begin{equation*}
	c_{n,m} = |\gamma_{n,m}(f + u)|^{2}\;,
	\qquad n\in\ZN\;,\ m=1,\dots,K^{2}
\end{equation*}
up to a global unitary factor.
\end{theorem}

\begin{remark}
The restriction on the norm in the definition of the signal space $\mathcal{S}^{p}$ requires only that we need to know an upper bound on the signal norm.
By the Theorem of Plancherel-P{\'o}lya \eqref{equ:PlPol}, this is equivalent to a restriction on the peak value of the signal.
This knowledge is necessary to choose an appropriated function $u$.
\end{remark}

The proof of Theorem~\ref{thm:JFAA_Binfty} is based on the following two facts:

\begin{enumerate}
\item
Let $\Lambda = \{\lambda_{n} = \xi_{n} + \I\eta_{n} \}_{n\in\ZN}$ be the zero set of an arbitrary sine-type function.
If one changes the imaginary parts of every $\lambda_{n}$, the resulting sequence is again the zero set of a sine-type function \cite{Levin_Perturbations}.

\item
Fix $1 \leq p \leq \infty$ and $\beta < \pi$.
Then to every $H_{u} > 0$ there exists a sine-type function $u$ such that for every $f \in \B^{p}_{\beta\pi}$ with $\|f\|_{\B^{p}_{\beta\pi}} \leq A_{\mathrm{max}}$, the function $v = f + u$ satisfies $|v(\xi + \I\eta)| > 0$ for all $\xi\in\RN$ and all $|\eta| > H_{u}$ \cite{PYB_JFAA14}.
So all zeros of $v = f + u \in \B^{\infty}_{\pi}$ are concentrated in a strip parallel to the real axis.

\end{enumerate}

So based on these two observations, we can choose $\Lambda = \{\lambda_{n} = \xi_{n} + \I\eta_{n}\}$ such that the measurement functionals satisfy the recovery condition. Then we choose an arbitrary $H_{u} > 0$ and increase (if necessary) all $\eta_{n}$ such that $|\eta_{n}| > H_{u}$ for all $n\in \ZN$. The resulting sequence will still satisfy the recovery condition. Then we choose $u$ such that the zeros of all function $v = f + u$ with $f \in \B^{p}_{\beta\pi}$ lie close to the real axis. In this way, we can achieve that the functions $v = f + u$ will definitely have no zero on the set $\Lambda$ and we can recover $v$ from the measurements $|\gamma_{n}(v)|^{2}$. Since $u$ is known, we can finally determine $f$.

The second step of the recovery algorithm consist in the interpolation of $v$ from the samples $\{v(\lambda_{n}\}_{n}$.
Since $v \in \B^{\infty}_{\pi}$, we necessarily need to apply the results for the sampling series on $\B^{\infty}_{\pi}$ as discussed in Section~\ref{sec:ConvOverSamp}.
In particular, it follows from Theorem~\ref{thm:ConvergBinfty} that we necessarily need oversampling, i.e. Theorem~\ref{thm:JFAA_Binfty} does not hold for functions in $\B^{p}_{\pi}$.

\begin{acknowledgement}
We thank Joachim Hagenauer and Sergio Verd\'u for drawing our attention to \cite{Butzer_RaabesWork_AA11} and for related discussions, and Ullrich M{\"o}nich for carefully reading the manuscript and for helpful comments.
The first author thanks the referees of the German Research Foundation (DFG) grant BO~1734/13-2 for highlighting the importance of understanding the strong divergence behavior addressed in Section~\ref{sec:StrongDiv} of this paper. He also likes to thank Rudolf Mathar for his insistence in several conversations on the significance of these questions.

The authors gratefully acknowledge support by the DFG through grants BO~1734/22-1 and PO~1347/2-1.
\end{acknowledgement}

% ==============================
% ========== Appendix ==========
% ==============================
\section*{Appendix}
%\addcontentsline{toc}{section}{Appendix}
%%
%%
%When placed at the end of a chapter or contribution (as opposed to at the end of the book), the numbering of tables, figures, and equations in the appendix section continues on from that in the main text. Hence please \textit{do not} use the \verb|appendix| command when writing an appendix at the end of your chapter or contribution. If there is only one the appendix is designated ``Appendix'', or ``Appendix 1'', or ``Appendix 2'', etc. if there is more than one.

This appendix provides a short proof of Theorem~\ref{thm:DivSets} in Section~\ref{sec:StrongDiv}

\begin{proof}[Theorem~\ref{thm:DivSets}]
1. First, we prove the statement for the sets \eqref{equ:UnionD}.
To this end, let $g \in \mathcal{X}$ and $\epsilon > 0$ be arbitrary. 
For all $M, N_{0} \in \NN$, we have to show that there exists a functions $f_{*}$ in the set \eqref{equ:UnionD} such that $\|g - f_{*}\|_{\mathcal{X}} < \epsilon$.
Since $\mathcal{X}_{0}$ is a dense subset of $\mathcal{X}$, there exists a $q \in \mathcal{X}_{0}$ such that $\|g - q\|_{\mathcal{X}} < \epsilon/2$.
Therewith, we define
$f_{*} := q + \tfrac{\epsilon}{2}\, f_{0}$
with $f_{0} \in \mathcal{D}_{weak}$ and with $\|f_{0}\|_{\mathcal{X}} = 1$.
Then we get
\begin{equation*}
	\|g - f_{*}\|_{\mathcal{X}}
	\leq \|g - q \|_{\mathcal{X}} + \tfrac{\epsilon}{2} \|f_{0}\|_{\mathcal{X}}
	< \tfrac{\epsilon}{2} + \tfrac{\epsilon}{2} = \epsilon\;.
\end{equation*}
Let $M, N_{0} \in \NN$ be arbitrary. We still have to show that $f_{*}$ is contained in the set \eqref{equ:UnionD}.
To this end, we observe that for every $N\in\NN$
\begin{equation*}
	\| \Op{T}_{N} f_{*} \|_{\mathcal{Y}}
	= \|\Op{T}_{N} q + \tfrac{\epsilon}{2}\Op{T}_{N} f_{0} \|_{\mathcal{Y}}
	\geq \tfrac{\epsilon}{2} \|\Op{T}_{N} f_{0} \|_{\mathcal{Y}} - \|\Op{T}_{N} q\|_{\mathcal{Y}}\;.
\end{equation*}
Since $q \in \mathcal{X}_{0}$, \eqref{equ:ConvOnDenseSubSet} implies that there is an $N_{1} \geq N_{0}$ such that
$1 \geq \| \Op{T}_{N}q - q \|_{\mathcal{Y}} \geq \| \Op{T}_{N}q \|_{\mathcal{Y}} - \|q\|_{\mathcal{Y}}$ for all $N\geq N_{1}$.
Consequently
\begin{equation*}
	\|\Op{T}_{N} q\|_{\mathcal{Y}} \leq 1 + \|q\|_{\mathcal{Y}} \leq 1 + C_{0}\|q\|_{\mathcal{X}}
	\quad\text{for all}\ N \geq N_{1}\;,
\end{equation*}
using for the last inequality that $\mathcal{X}$ is continuously embedded in $\mathcal{Y}$ with a certain constant $C_{0} < \infty$.
Combining the last two inequalities, we get
$\| \Op{T}_{N} f_{*} \|_{\mathcal{Y}} \geq \tfrac{\epsilon}{2} \|\Op{T}_{N} f_{0}\|_{\mathcal{Y}} - 1 - C_{0}\|q\|_{\mathcal{X}}$
for all $N \geq N_{1}$.
Since $f_{0} \in \mathcal{D}_{\mathrm{weak}}$ there exits an $N_{2} \geq N_{1}$ such that
\begin{equation*}
	\| \Op{T}_{N_{2}} f_{*} \|_{\mathcal{Y}} 
	\geq \tfrac{\epsilon}{2} \| \Op{T}_{N_{2}} f_{0} \| - 1 - C_{0} \|q\|_{\mathcal{X}} > M
\end{equation*}
which shows that 
$f_{*} \in D(M,N_{2}) \subset \bigcup_{N \geq N_{0}} D(M,N)$.
Thus the sets \eqref{equ:UnionD} are dense in $\mathcal{X}$ and it remains to show that these sets are open.
To this end, let $M, N \in \NN$ and $f_{*} \in D(M,N)$ be arbitrary, i.e. $\|\Op{T}_{N} f_{*}\|_{\mathcal{Y}} > M$.
Since $\Op{T}_{N}$ is a continuous linear operator $\mathcal{X} \to \mathcal{Y}$, there exists a $\delta > 0$ and a neighborhood 
\begin{equation*}
	U_{\delta}(f_{*}) = \{ f \in \mathcal{X} : \|f - f_{*}\|_{\mathcal{X}} < \delta\}
\end{equation*}
of $f_{*}$ such that $\|\Op{T}_{N} f\|_{\mathcal{Y}} > M$ for all $f \in U_{\delta}$.
Thus $D(M,N)$ is open for all $M,N \in \NN$ and since the union of (countable many) open sets is again open, the sets \eqref{equ:UnionD} are also open.

2. We prove \eqref{equ:LemDweak}. By the definition of the $\limsup$ operation, the set $\mathcal{D}_{\mathrm{weak}}$ can be written as
\begin{equation*}
	\mathcal{D}_{\mathrm{weak}} = \Big\{ f \in \mathcal{X}\ :\ \lim_{N_{0}\to\infty} \sup_{N \geq N_{0}} \|\Op{T}_{N} f\|_{\mathcal{Y}} = \infty \Big\}
\end{equation*}
and we note that for every fixed $f\in \mathcal{X}$ the sequence $\{ \sup_{N \geq N_{0}} \|\Op{T}_{N} f\|_{\mathcal{Y}} \}^{\infty}_{N_{0}=1}$ is monotone decreasing.
Assume that $f \in \mathcal{D}_{\mathrm{weak}}$ and choose $M \in \NN$ arbitrary.
Then, by the above definition of $\mathcal{D}_{\mathrm{weak}}$, it follows that for arbitrary $N_{0}$ there exists an $N \geq N_{0}$ such that $\|\Op{T}_{N} f\|_{\mathcal{Y}} > M$, i.e.
$f \in \bigcup_{N\geq N_{0}} D(M,N)$, and since this holds for all $M, N \in \NN$ we have
\begin{equation*}
	f \in \bigcap^{\infty}_{M=1}\ \bigcap^{\infty}_{N_{0} = 1}\ \bigcup^{\infty}_{N = N_{0}}\ D(M,N)
\end{equation*}
which shows that $\mathcal{D}_{\mathrm{weak}} \subset \bigcap^{\infty}_{M=1}\ \bigcap^{\infty}_{N_{0} = 1}\ \bigcup^{\infty}_{N = N_{0}}\ D(M,N)$.
Conversely, assume that $f \in \bigcap^{\infty}_{M=1}\ \bigcap^{\infty}_{N_{0} = 1}\ \bigcup^{\infty}_{N = N_{0}}\ D(M,N)$.
Then to every arbitrary $M \in \NN$ and $N_{0} \in \NN$ there exists an $N > N_{0}$ such that $f \in D(M,N)$, i.e. that $\|\Op{T}_{N} f\|_{\mathcal{Y}} > M$. Thus $f \in \mathcal{D}_{\mathrm{weak}}$.

Finally, we prove \eqref{equ:LemDstrong}. Assume first that $f \in \mathcal{D}_{\mathrm{strong}}$.
Then to every $M \in \NN$ there exists an $N_{0} = N_{0}(M)$ such that $\|\Op{T}_{N} f\|_{\mathcal{Y}} > M$ for all $N \geq N_{0}$. In other words
\begin{equation*}
	f \in \bigcap^{\infty}_{N=N_{0}(M)} D(M,N) \subset \bigcup^{\infty}_{N_0=1} \bigcap^{\infty}_{N=N_{0}} D(M,N)
	\qquad\text{for every}\ M\in\NN\;.
\end{equation*}
which shows that $f \in \bigcap^{\infty}_{M=1}\ \bigcup^{\infty}_{N_{0} = 1}\ \bigcap^{\infty}_{N = N_{0}}\ D(M,N)$.
Conversely, assume that $f \in \bigcap^{\infty}_{M=1}\ \bigcup^{\infty}_{N_{0} = 1}\ \bigcap^{\infty}_{N = N_{0}}\ D(M,N)$.
This means that for any arbitrary $M\in\NN$ the function $f$ belongs to $\bigcup^{\infty}_{N_{0} = 1}\ \bigcap^{\infty}_{N = N_{0}}\ D(M,N)$, i.e. there exists an $N_{0}$ such that
\begin{equation*}
	f \in \bigcap^{\infty}_{N=N_{0}} D(M,N)
	\qquad\text{i.e.}\qquad
	\|\Op{T}_{N} f\|_{\mathcal{Y}} > M 
	\quad\text{for all}\ N \geq N_{0}\;.
\end{equation*}
Thus $\lim_{N\to\infty} \|\Op{T}_{N} f\|_{\mathcal{Y}} = \infty$, i.e. $f \in \mathcal{D}_{\mathrm{strong}}$.
\qed\end{proof}

%\input{referenc}

% ================================================
% ================= Bibliographie ================
% ================================================
\renewcommand\refname{}\vspace{-2.5\baselineskip}
%\bibliographystyle{abbrvnat}
%\bibliographystyle{abbrv}

%\bibliography{bib}%,../publications}

%\printindex

\end{document}